\documentclass[manuscript]{acmart}
\AtBeginDocument{%
  }

\setcopyright{acmlicensed}
\copyrightyear{2026}
\acmYear{2026}
\acmDOI{XXXXXXX.XXXXXXX}

\usepackage{pifont}
\usepackage{soul}
\usepackage{array}
\usepackage{graphicx}
\usepackage{tabularx}
\usepackage{comment}
\usepackage{multirow}
\usepackage[dvipsnames,table]{xcolor}
\usepackage{colortbl}
\usepackage{longtable}
\usepackage{url}
\usepackage{adjustbox} 
\definecolor{LightCyan}{rgb}{0.88,1,1}

\begin{document}

\title{Towards Secure MLOps: Surveying Attacks, Mitigation Strategies, and Research Challenges}

\author{Raj Patel}
\email{rpatel38@ua.edu}
\authornotemark[1]
\affiliation{%
  \institution{The University of Alabama}
  \city{Tuscaloosa}
  \state{Alabama}
  \country{USA}}

\author{Himanshu Tripathi}
\affiliation{%
  \institution{Mississippi State University}
  \city{Mississippi State}
  \country{USA}}
\email{ht557@msstate.edu}

\author{Jasper Stone}
\affiliation{%
  \institution{Mississippi State University}
  \city{Mississippi State}
  \country{USA}}
\email{jws819@msstate.edu}

\author{Noorbakhsh Amiri Golilarz}
\email{noor.amiri@ua.edu}
\affiliation{%
  \institution{The University of Alabama}
  \city{Tuscaloosa}
  \state{Alabama}
  \country{USA}
}

\author{Sudip Mittal}
\email{sudip.mittal@ua.edu}
\affiliation{%
  \institution{The University of Alabama}
  \city{Tuscaloosa}
  \state{Alabama}
  \country{USA}
}

\author{Shahram Rahimi}
\email{srahimi1@ua.edu}
\affiliation{%
  \institution{The University of Alabama}
  \city{Tuscaloosa}
  \state{Alabama}
  \country{USA}
}

\author{Vini Chaudhary}
\email{vchaudhary@cse.msstate.edu}
\affiliation{%
  \institution{Mississippi State University}
  \city{Mississippi State}
  \country{USA}}


\renewcommand{\shortauthors}{Patel et al.}

\begin{abstract}
    The rapid adoption of machine learning (ML) technologies has driven organizations across diverse sectors to seek efficient and reliable methods to accelerate model development-to-deployment. Machine Learning Operations (MLOps) has emerged as an integrative approach addressing these requirements by unifying relevant roles and streamlining ML workflows. As the MLOps market continues to grow, securing these pipelines has become increasingly critical. However, the unified nature of MLOps ecosystem introduces vulnerabilities, making them susceptible to adversarial attacks where a single misconfiguration can lead to compromised credentials, severe financial losses, damaged public trust, and the poisoning of training data. Our paper presents a systematic application of the MITRE ATLAS (Adversarial Threat Landscape for Artificial-Intelligence Systems) framework, supplemented by reviews of white and grey literature, to systematically assess attacks across different phases of the MLOps ecosystem. We begin by reviewing prior work in this domain, then present our taxonomy and introduce a threat model that captures attackers with different knowledge and capabilities. We then present a structured taxonomy of attack techniques explicitly mapped to corresponding phases of the MLOps ecosystem, supported by examples drawn from red-teaming exercises and real-world incidents. This is followed by a taxonomy of mitigation strategies aligned with these attack categories, offering actionable early-stage defenses to strengthen the security of MLOps ecosystem. Given the gradual evolution and adoption of MLOps, we further highlight key research gaps that require immediate attention. Our work emphasizes the importance of implementing robust security protocols from the outset, empowering practitioners to safeguard MLOps ecosystem against evolving cyber attacks. 
\end{abstract}




\maketitle

\section{Introduction} \label{sec:introduction}
Machine Learning (ML) technologies have fueled remarkable growth across diverse industries, prompting organizations to swiftly transition models from development into production to secure competitive advantages. To manage this acceleration, companies are increasingly adopting Machine Learning Operations (MLOps) lifecycle framework \cite{grand_view_research}. First introduced in 2015 \cite{mlops-coin}, MLOps is defined as ``a set of practices designed to create an assembly line for building and running machine learning models. It helps companies automate tasks and deploy models quickly, ensuring everyone involved (data scientists, engineers, IT) can cooperate smoothly and monitor and improve models for better accuracy and performance'' \cite{IBM-mlops}. The MLOps market reached \$2.2B in 2024 \cite{grand_view_research}, with projections indicating a 43\% expansion over the next five years \cite{stone2025mlops}. Industry adoption is evidenced by full MLOps implementations at major companies including Netflix, Uber, DoorDash, and Lush \cite{MLOpsAdoption}. Major cloud providers and vendors now offer integrated MLOps platforms supporting data ingestion, training, deployment, and monitoring at enterprise scale \cite{ibm_xforce_mlops}. These developments collectively demonstrate that MLOps has become a central coordinating layer in modern AI-enabled infrastructures.

The strength of MLOps arises from combining the iterative nature of ML with the Continuous Integration and Continuous Delivery (CI/CD) practices established by DevOps. DevOps refers to ``a set of practices for automating the processes between software development and information technology operations teams so that they can build, test, and release software faster and more reliably. The goal is to shorten the systems development life cycle and improve reliability while delivering features, fixes, and updates frequently in close alignment with business objectives” \cite{nist-devops}. Integrating DevOps methodologies with ML workflows results in robust and flexible MLOps ecosystem capable of rapidly responding to evolving business and security requirements.

Although MLOps offers substantial benefits, the pressure to accelerate model deployment for competitive gain often results in insufficient attention to security, increasing exposure to vulnerabilities across the MLOps ecosystem. Research by IBM X-Force demonstrates that widely used enterprise MLOps platforms, including Azure Machine Learning, BigML, and Google Cloud Vertex AI, are vulnerable to attacks by adversaries with valid credentials, enabling training data poisoning, model and training data theft, and exploitation of production pipelines and connected data lakes \cite{ibm_xforce_mlops}, exposing critical Machine Learning Security Operations (MLSecOps) concerns \cite{calefato_Sec_Risk_MLOps}. With the MLOps market projected to grow 43\% over the next five years \cite{grand_view_research,stone2025mlops}, securing these pipelines has become critical. High-profile incidents such as the ShadowRay vulnerability \cite{ShadowRay}, where insufficient authentication control mechanisms allowed attackers to seize compute resources and exfiltrate more than \$1 billion worth of data in September 2023, underline the severity of these risks. Similarly, the Arbitrary Code Execution incident involving Google Colab \cite{google_colab_code_execution} highlighted the dangers of malicious scripts hidden in shared Jupyter Notebooks, emphasizing the urgent need for rigorous code review processes. DeepSeek's V3 model \cite{DeepSeek} allegedly resulted from repeated unauthorized distillation queries to OpenAI's models \cite{DeepSeek-distillation}, demonstrating intellectual property theft risks. Additionally, research conducted by SpiderSilk \cite{ClearviewAI} demonstrated how a single misconfiguration compromised critical credentials, leading to unauthorized access to Clearview AI's sensitive code repositories. Given that Clearview AI's facial recognition tools are used in matching identities using publicly available images, the implications of such a breach are especially concerning. These examples collectively highlight how security lapses in the MLOps ecosystem can lead to infiltration, data theft, operational disruption, and substantial reputational damage.

Traditional security measures, such as deploying standard malware or virus detection tools within MLOps ecosystem, are insufficient to counter these evolving attacks. Adversaries continuously adapt and refine their tactics, employing defense-aware strategies capable of evading traditional security measures. Skylight’s findings \cite{cylance2019} demonstrated how adversarial inputs could evade current malware detection systems. In 2020, Palo Alto Networks' Security AI research team \cite{DGA_detection} showcased how generic domain name mutation techniques could defeat Convolutional Neural Network (CNN)-based DGA detection modules. Such mutations evade most ML-based detection systems, underscoring the importance of thorough robustness evaluation before deployment. These attacks often leverage Open Source Intelligence Gathering (OSINT) \cite{OSINT} to maximize infiltration rates while minimizing detection and attribution. Such examples underscore the need to embed security deeply within the MLOps ecosystem, addressing threats at every stage of the ML lifecycle.

In light of these gaps, our study is guided by the following research questions:
\begin{itemize}
    \item \textbf{RQ1:} What resources exist that systematically catalog AI-related attacks and attacker techniques, and how can they be leveraged to comprehensively map vulnerabilities within the MLOps ecosystem?
    \item \textbf{RQ2:} What is the severity of the threats presented to MLOps ecosystem, and how many of these threats have been observed in real-world incidents (or case studies), as opposed to remaining theoretical in nature?
    \item \textbf{RQ3:} How well do existing adversarial frameworks align their proposed defenses and mitigations with the identified attack techniques, and to what extent do they provide comprehensive coverage of this attack surface?
    \item \textbf{RQ4:} What notable gaps persist, and what further research is necessary to enhance security across the MLOps ecosystem?
\end{itemize}

To address these research questions and challenges, this paper provides the first comprehensive mapping of MLOps-centric attacks using the MITRE ATLAS (Adversarial Threat Landscape for Artificial Intelligence Systems) framework \cite{mitreatlas}, an exhaustive and continuously updated catalog of AI-centric attack lifecycles. Additionally, our analysis integrates established frameworks such as MITRE ATT\&CK \cite{mitre_attack} and the emerging OWASP (Open Worldwide Application Security Project) Top 10 for Large Language Models \cite{owasp} with extensive white and grey literature, including security research papers, industry reports, documented incidents, and red teaming exercises. This multi-source approach enables us to identify key security risks and validate attack techniques through real-world evidence. By integrating these resources, we enhance our attack assessment and address existing gaps within the MITRE ATLAS framework. Leveraging the ATLAS framework enables organizations to perform structured evaluations of attack vectors, uncover vulnerabilities, and integrate security measures throughout the MLOps ecosystem. Our extensive review of mitigation strategies further assists organizations in protecting mission-critical assets, reinforcing resilience against continuously evolving attacks. The major contributions of this paper are as follows:

\begin{itemize}
    \item We highlight how adversaries exploit vulnerabilities at each phase of the MLOps ecosystem using the ATLAS framework together with relevant white and grey literature to identify potential attack vectors.
    \item We review a comprehensive taxonomy of MLOps-centric attacks and validate their applicability through examples drawn from real-world incidents and/or red-team exercises documented in the ATLAS framework.
    \item We present a structured taxonomy of mitigation strategies to enhance the detection and prevention of adversarial attacks on MLOps ecosystem.
    \item We discuss challenges and provide research recommendations to enhance the security and robustness of the MLOps ecosystem, ensuring resilience against emerging sophisticated cyberattacks.
\end{itemize}

The remainder of this paper presents a comprehensive analysis of adversarial risks inherent to MLOps ecosystems. Section II introduces the MLOps ecosystem, detailing foundational concepts, operational taxonomy, and real-world implementations. Section III reviews prior work in this domain and explains how our study is positioned, then sets out the survey taxonomy and defines a threat model that captures attackers with different motivations, knowledge, and capabilities. Section IV leverages the MITRE ATLAS framework to review the evolving attack landscape, illustrating the impact of these threats through empirical evidence from real-world incidents and red-teaming exercises. Section V presents mitigation strategies addressing systemic vulnerabilities within the MLOps ecosystem. Finally, Section VI outlines ongoing research challenges in securing MLOps and identifies opportunities to strengthen defenses through integration of emerging security technologies.

\section{MLOps Overview} \label{sec:MLOpsOverview}
The concept of Machine Learning Operations (MLOps) emerged in 2015 \cite{mlops-coin} to bridge the gap between isolated model development to practical deployment. MLOps enhances automation, communication, and monitoring within ML workflow that are inspired by DevOps principles, aligning them with organization's objective and creating tangible value \cite{tolpegin2020data}. Large Language Model Operations (LLMOps) is a specialized subset of MLOps where replacing model, data, and evaluation blocks with LLM oriented versions preserves the pipeline structure \cite{llmops,stone2025mlops}.

\begin{figure}[!htb]
    \centering
    \includegraphics[width=\textwidth]{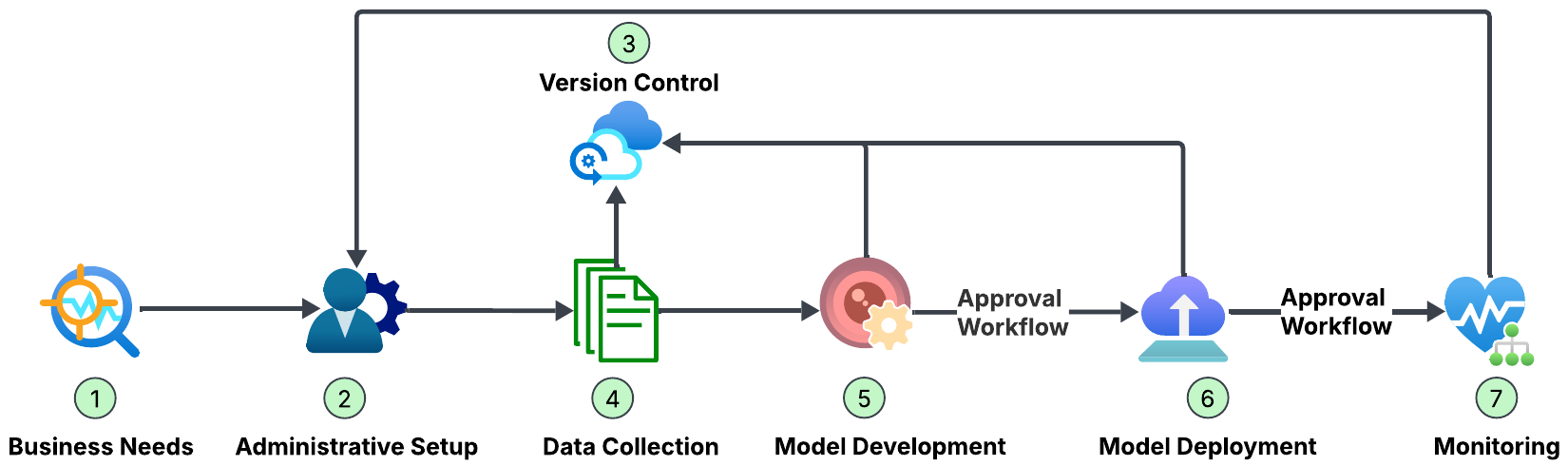}
    \caption{The MLOps ecosystem begins by identifying business needs, followed by administrative setup, model development, and deployment. Continuous monitoring then provides essential feedback, enabling iterative improvements.}
    \Description{A high-level consolidated architecture of MLOps ecosystem is presented.}
    \label{fig:MLOPS}
\end{figure}

Fig.~\ref{fig:MLOPS} provides a visual summary of the MLOps ecosystem, which is explored in greater depth in \cite{stone2025mlops}. Here, we outline the key phases most relevant to the security considerations discussed in the following sections. The MLOps process begins with the clear identification of business needs and the establishment of specific Objectives and Key Results (OKRs) that guide decision-making and operational priorities \cite{testi2022mlops} \cite{Eken_Multivocal_Review_MLOps}. Effective machine learning models can directly influence commercial outcomes, and a scalable MLOps ecosystem enables the continuous training and deployment of projects as they move from proof-of-concept to production scale \cite{makinen2021needs}.

Following business alignment, data collection uses many methods including web scraping, APIs, IoT sensors, and crowdsourcing \cite{yerlikaya2022data, tolpegin2020data, makinen2021needs}. Augmentation and transfer learning can enhance the data that is readily available in situations where information is limited \cite{kreuzberger2023machine, zhang2022conceptualizing}. Data preparation typically begins with cleaning procedures that address missing values, errors, and duplicates. This is followed by normalization techniques to ensure data consistency and enhance algorithmic efficiency \cite{kumara2022requirements, yerlikaya2022data}. Dataset reliability is improved through robust versioning and seamless data integration, which in turn support effective feature engineering and exploratory data analysis (EDA) \cite{makinen2021needs}.

A robust administrative framework underpins scalable MLOps infrastructure by supporting key operations such as resource provisioning, user role management, and system monitoring \cite{stone2025mlops}. It ensures access to necessary computational resources, including CPUs, GPUs, and TPUs, for efficient training and deployment. Monitoring tools track system health and anomalies to maintain model consistency and integrity. Additionally, well-defined roles, permissions, and Identity and Access Management (IAM) controls \cite{amazon-iam} promote secure collaboration and safeguard against unauthorized access.

The model development phase focuses on transforming data into actionable insights through robust experimentation and iterative refinement. Key tasks include data formatting, exploratory data analysis (EDA), feature engineering, and algorithm selection. Using statistical and visual techniques, EDA uncovers patterns, relationships, and anomalies that guide effective feature selection \cite{John2021, Lima2022}. Feature engineering, through methods such as dimensionality reduction and polynomial expansion, refines raw inputs to improve model accuracy and robustness \cite{makinen2021needs, kreuzberger2023machine}. Algorithm selection, which includes methods such as decision trees, linear regression, and neural networks, is informed by both data characteristics and organizational needs \cite{testi2022mlops, kreuzberger2023machine}. Rigorous evaluation metrics, including precision, F1-score, ROC-AUC, and MSE, along with considerations for model interpretability, help ensure alignment with strategic goals and build stakeholder trust \cite{zhang2022conceptualizing, testi2022mlops, kreuzberger2023machine}. To mitigate overfitting and improve inference, training procedures incorporate techniques such as early stopping, adaptive learning rates, and cross-validation \cite{zhang2022conceptualizing, Lima2022}.

Version control and experiment tracking allow teams to document and monitor changes to data, features, hyperparameters, and source code, promoting consistency and reproducibility. These systems also provide rollback capabilities in case of performance degradation, helping maintain the stability and reliability of the MLOps workflow \cite{Lima2022, kreuzberger2023machine}. Continuous Integration (CI) automates the testing and validation of new code and model updates, ensuring compatibility and minimizing deployment risks \cite{9723793, google-mlops}.

The deployment phase emphasizes rigorous testing and validation under conditions that closely mirror the production environment, ensuring both reliability and performance. Quality assurance procedures such as batch processing and real-time inference validations are conducted to verify that models meet predefined accuracy and latency requirements \cite{9355312}. Responsible AI practices are also incorporated at this stage to mitigate bias and promote fairness. Before release into production, models must pass through a gated approval workflow that includes a combination of automated performance evaluations and manual reviews to ensure compliance with robustness and security standards \cite{stone2025mlops}. Once approved, models are deployed for end-user access and configured to support diverse inference scenarios. Benchmarking tools, including MLPerf \cite{mlperf}, are used to assess production readiness and verify that inference quality is maintained in the target environment \cite{9138989}.

Continuous monitoring is essential in complex multi-cloud environments, enabling performance tracking across testing, staging, and production stages to ensure timely detection and resolution of deviations \cite{6740239}. This comprehensive approach facilitates sustained alignment with evolving business needs.

In summary, the comprehensive MLOps ecosystem effectively manages the machine learning lifecycle, integrating continuous development, deployment, and monitoring. By prioritizing alignment with business objectives and responsible AI practices, organizations can rigorously evaluate and continuously monitor the performance of their machine learning systems. These efforts contribute to operational efficiency, ethical model behavior, sustained reliability, and overall trustworthiness. However, despite these advancements, MLOps ecosystem remain vulnerable to security threats. Ensuring long-term dependability and stakeholder trust requires ongoing research and integration of robust, context-aware security.

\begin{figure}[ht!]
    \includegraphics[scale=0.55]{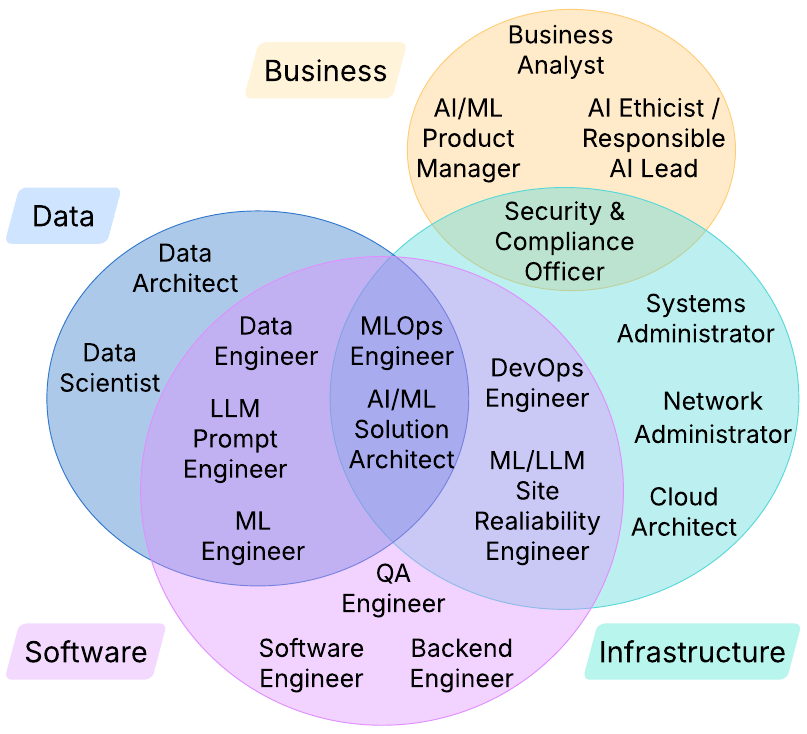}
    \caption{Venn diagram of potential operational roles contributing to secure MLOps ecosystem development.}
    \Description{Venn diagram depicting overlapping of roles and stakeholders involved in MLOps security.}
    \label{fig:intended-audience}
\end{figure}

\section{Survey Approach And Taxonomy} \label{sec:survey_approach_taxonomy}
The preceding section examined the MLOps ecosystem and described how tools, roles, and software components interact to automate the machine learning lifecycle. This automation accelerates development and operational reliability but also inherits security exposure at each stage of the lifecycle, especially when common standards are missing and organizations combine heterogeneous platforms and tools \cite{Eken_Multivocal_Review_MLOps}. In this section, we focus on how these security exposures manifest when viewing MLOps as an end-to-end socio-technical system rather than as a group of isolated components. To address RQ1, we draw on peer-reviewed research, technical blogs, industry reports, and public frameworks such as MITRE ATLAS and MITRE ATT\&CK, to connect concrete attacker tactics and techniques to the main stages of the MLOps lifecycle \cite{mitreatlas, mitre_attack}. To our knowledge, this is the first survey to systematically map attacker techniques to the full MLOps and LLMOps lifecycle, and to use this mapping as the foundation for both a security taxonomy and a threat model. The remainder of this section explains how our work is situated in the literature, how we structure the MLOps ecosystem into three taxonomical families, and how we use this structure to define a threat model that reflects attacker behavior and operational concerns.

\subsection{Survey Landscape and Positioning}
The machine learning security research community has typically studied attacks and defenses by focusing on individual attack families such as poisoning attacks, privacy attacks, and evasion attacks, with an emphasis on model or data behavior in controlled settings. These studies provide insight into how specific techniques work and how models can fail. However, they rarely analyze these attacks within the context of the complete MLOps lifecycle, which spans design, development, deployment, and operations. As a result, most surveys consider each phase of the lifecycle as independent, even though decisions made during early design stages directly shape downstream risks. Only a limited number of works consider an interconnected perspective that takes into account how data pipelines, training workflows, deployment stacks, and monitoring systems share common infrastructure. In addition, most prior surveys approach the subject from a research or platform engineering viewpoint and give limited attention to the needs of end users and operators who are responsible for maintaining secure and reliable systems over time. In this subsection, we review representative surveys and frameworks on secure MLOps and SecMLOps, and clarify how our work is positioned both from a research perspective and from an operational perspective. This review motivates our focus on security across the entire lifecycle and prepares the ground for the taxonomy that follows.

Calefato et al. \cite{calefato_Sec_Risk_MLOps} synthesize academic studies and practitioner reports to create a catalog of risks and best practices for machine learning operations, covering access control, network protection, secure deployment, monitoring, privacy, supply chain integrity, and organizational culture. They identify recurring risks including unauthorized access, data exfiltration, data poisoning, model theft, code injection, unpatched dependencies, weak monitoring, and limited security awareness. Their recommendations include managed identities, strict access control with full logging, network isolation, end-to-end encryption, comprehensive input and output validation, provenance tracking, anomaly detection, adversarial training, careful dependency management, and incident response planning. Framing these practices as part of the MLSecOps discipline, their work emphasizes the need for stronger protection of data and models together with continuous monitoring and incident readiness. However, their catalog does not explicitly map attacker tactics to lifecycle stages or to public knowledge bases such as ATLAS. Our survey builds on their foundation by using ATLAS to organize risks and defenses according to specific tactics and techniques observed in real MLOps deployments.

Ahmad et al. \cite{Ahmad_MLOps_Enabled_Security} extend the discussion to operational technology environments and argue that traditional IT controls are insufficient where machine learning systems impact physical processes and safety-critical applications. They call for comprehensive risk management across data, models, pipelines, infrastructure, personnel, and third-party dependencies, supported by data encryption, rigorous access control, anonymization, secure model versioning and storage, secured CI/CD pipelines, automated code scanning, container-based deployment, and continuous monitoring with audits. Zhang et al. \cite{zhang2022conceptualizing} introduce the SecMLOps paradigm, embedding security throughout the lifecycle using a People, Processes, Technology, Governance, and Compliance lens. They define clear security responsibilities for stakeholders including business leads, data engineers, data scientists, DevOps engineers, and MLOps engineers. They also organize security activities such as threat modeling, policy definition, model security evaluation, and incident response into design, experimentation, and production stages, while considering how these intersect with fairness, explainability, reliability, safety, and sustainability.

Panchumarthi \cite{panchumarthi2025devsecmlops} proposes a DevSecMLOps framework that adapts DevSecOps practices for machine learning by covering data ingestion with automated detection and anonymization, model training with dependency scanning and reproducibility checks, model packaging with container scanning and software bill of materials generation, deployment using Zero Trust controls and policy as code, and operations with drift detection and automated incident response. Case studies in that work show improvements in vulnerability detection, lower false positive rates, better policy compliance, and faster drift detection. The framework also generates audit-ready artifacts to support regulatory requirements including GDPR, HIPAA, SOC 2, ISO 27001, and PCI DSS. While these works outline key elements of secure MLOps and SecMLOps, they do not provide a comprehensive mapping between attacker tactics, lifecycle stages, and operator-facing responsibilities across the ecosystem.

Our survey addresses these gaps by taking a lifecycle-oriented and operations-focused view that treats MLOps and LLMOps as interconnected systems rather than isolated components. We use ATLAS to connect attacker tactics to specific decisions and components across the lifecycle, interpreting these connections through two main perspectives: one for researchers who study attacks and defenses, and one for security and operations teams who deploy and maintain these systems as shown in Fig. \ref{fig:intended-audience}. From the research perspective, our mapping offers a structured way to relate attack families such as poisoning, privacy, and evasion to the broader MLOps context and to highlight where existing techniques provide coverage and where gaps remain. From the operational perspective, our taxonomy tracks how design, platform configuration, development, deployment, and monitoring strategies influence exposure to specific ATLAS tactics. This dual perspective positions our survey relative to prior work and focuses attention on lifecycle stages that are most critical for operators and end users. The next subsection formalizes the taxonomy that underpins this approach and provides the structure used in later sections to analyze attacks, defenses, and attacker capabilities. By grounding this taxonomy in both the literature and operational practice, we aim to make it useful for both system designers and those responsible for running systems in real-world environments.

    \begin{figure*}
        \includegraphics[scale=0.461]{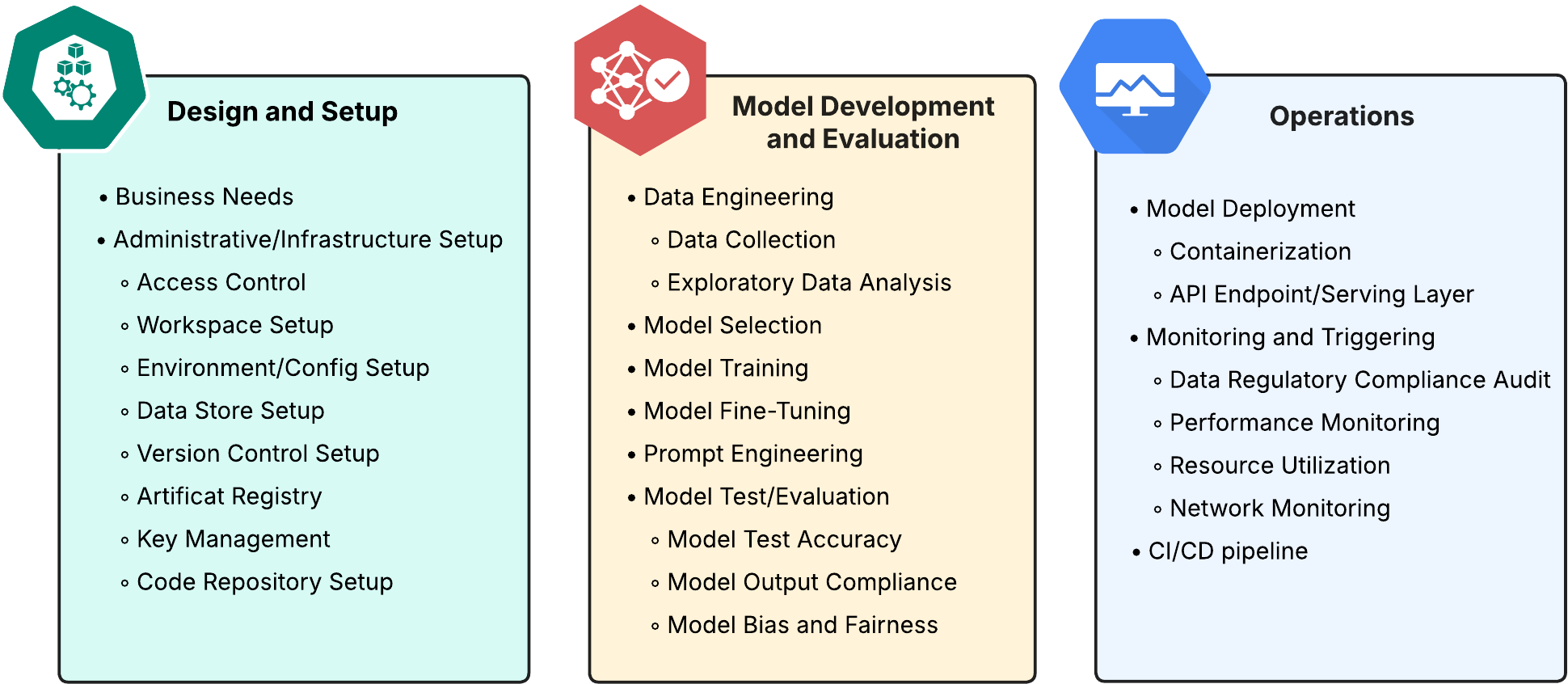}
        \caption{An overview of our proposed taxonomy. The MLOps lifecycle is structured into three distinct and non-overlapping families: Design and Setup, Model Development and Evaluation, and Operations. The bullet points listed under each family specify the corresponding phases within the lifecycle. This organizational structure is adapted from the principles in ml-ops \cite{mlops_principles_innoq} and is equally applicable to LLMOps.}
        \Description{Diagram grouping MITRE ATLAS tactics into columns with their respective attack techniques in the MLOps lifecycle.}
        \label{fig:MLOps_Survey_Taxonomy}
    \end{figure*}

\subsection{Survey Taxonomy} \label{sec:survey_taxonomy}
Guided by the landscape above, the MLOps and LLMOps ecosystem can be grouped into three families where security-relevant assets and attack surfaces are most concentrated: \textit{Design and Setup}, \textit{Model Development and Evaluation}, and \textit{Operations}. Instead of restating the full lifecycle, these families are treated as clusters of components and interfaces that are accessible to attackers and serve as key points for organizations to apply controls, defenses, and mitigations. Fig.~\ref{fig:MLOps_Survey_Taxonomy} illustrates this taxonomy and shows how attack paths can cross family boundaries.

\subsubsection{Design and Setup} \label{sec:design_and_setup}
\textit{Design and Setup} forms the control plane where organizations define resource access, environment structure, and trust relationships with external services. Core assets include identity and access control systems, workspace and project layouts, configuration and secrets management, infrastructure as code, version control systems, and artifact registries. For LLMOps, this also covers vector stores, prompt templates, safety policies, and tool connectors used in later stages. The main attack surface involves privileged accounts, configuration channels, and shared infrastructure. Attackers may target identity providers and policies to gain or escalate access, exploit misconfigured isolation to cross project boundaries, tamper with infrastructure as code and base images, or modify repositories, registries, and secrets stores. These design decisions directly influence which attacks become feasible in other families and determine which defenses and mitigations remain effective throughout the lifecycle.

\subsubsection{Model Development and Evaluation} \label{sec:model_development_and_evaluation}
\textit{Model Development and Evaluation} includes the assets and workflows that turn raw or curated data into trained models, prompts, and evaluation artifacts. Key components are data pipelines, labeling workflows, notebooks and integrated development environments, experiment tracking, training clusters, job schedulers, and model or prompt registries. For LLMOps, this extends to prompt design environments, libraries, and evaluation tools. The main attack surface is shaped by data flows and development interfaces. Data ingestion can be exploited for poisoning, labeling tools and human-in-the-loop processes may be used to bias labels or steal information, and notebooks and trackers often operate with broad permissions that allow data theft or backdoor insertion. Training jobs and pipelines may be misconfigured to swap datasets or reroute outputs, while model or prompt registries may be altered so that compromised artifacts are mistakenly approved for deployment.

\subsubsection{Operations} \label{sec:operations}
\textit{Operations} covers the runtime surfaces where deployed models and LLM configurations interact with users, external systems, and monitoring tools. Assets in this family include serving layers and API endpoints, inference jobs, online feature and context stores, monitoring and logging systems, safety and policy engines, routing and orchestration components, and CI/CD pipelines. In LLMOps, this also covers vector search services, tool execution frameworks, and gateways that proxy calls to external providers. Public or partner-facing endpoints can be exploited to send malicious inputs for prompt injection, model theft, or resource exhaustion. Weak authentication or opaque routing may lead to tenant data exposure. Feature stores, context stores, and vector indices can be poisoned or queried for sensitive data, while monitoring and logging systems may be manipulated to conceal or exaggerate incidents. CI/CD pipelines present a tightly connected surface where modified artifacts, policies, or configurations can be introduced as legitimate updates.

Across all three families, the taxonomy separates attack surfaces according to dominant assets and the specific interactions each surface exposes to attackers. \textit{Design and Setup} focuses on persistent control plane decisions, \textit{Model Development and Evaluation} centers on data and training workflows, and \textit{Operations} highlights runtime exposure and automated change. This structure is used in Section~\ref{sec:threat_model} to formalize the threat model and to organize the discussion of attacker techniques, defenses, and mitigations in later sections.

\subsection{Threat Model and Capabilities} \label{sec:threat_model}
Building on the taxonomy and incorporating ATLAS \cite{mitreatlas} and related references, this subsection develops a threat model from the perspective of the attacker, illustrating how adversaries develop motivation, gather knowledge, and acquire the capabilities necessary to compromise MLOps and LLMOps environments. The model conceptualizes the attacker’s progression through three interconnected phases: understanding the reasons for targeting machine learning systems (motivation and impact), identifying the assets and surfaces to target (intelligence and knowledge), and developing the means to execute attacks (capabilities and staging). These phases yield attacker profiles ranging from external black-box adversaries to privileged insiders, each engaging with the principal attack surfaces defined across the MLOps lifecycle.

\subsubsection{Motivation and Impact}
From an adversarial standpoint, MLOps and LLMOps environments present attractive targets because they control critical decisions and often lack comprehensive security controls \cite{Impact}. Attackers seek to influence the integrity, availability, confidentiality, and operational cost of machine learning systems, as well as to undermine the perceived trustworthiness of these platforms. Their motivations may be immediate and tactical, such as evading detection or reducing system reliability, or more strategic, involving persistent access and long-term manipulation. For example, attackers focused on short-term objectives may seek to bypass machine learning-based security controls by crafting adversarial inputs that cause misclassification \cite{CraftAdversarialData, Impact_EvadeMLModel}. Adversaries may also deny machine learning services through resource exhaustion attacks or by introducing chaff data, which degrades trust and may force organizations to revert to manual processes \cite{Impact_DenialofMLService, Impact_SpammingMLSystemwithChaffData}.

Economic drivers include cost abuse and intellectual property theft, where adversaries manipulate cloud-based machine learning workloads to increase operational expenses for victims or exfiltrate proprietary models and datasets to accelerate their own development \cite{Impact_CostHarvesting, Impact_ExternalHarmsMLIntellectualPropertyTheft}. Privacy-focused attackers exploit the capacity of models to memorize and leak sensitive or confidential information, leading to legal liability and reputational harm \cite{Impact, Impact_ExternalHarmsReputationalHarm}. Advanced attackers may pursue persistent objectives by poisoning training data or corrupting models, introducing hidden failure modes that persist beyond initial access \cite{Impact_ErodeDatasetIntegrity, Impact_ErodeMLModelIntegrity}. Documented incidents, including backdoor attacks with neural payloads, identity verification bypasses, and sensor-based adversarial manipulations, exemplify the diversity and severity of these threats \cite{deeppayload, BypassID.ME, cameraHijackChina, Impact, Impact_ExternalHarms}. Understanding these motivations is a prerequisite for analyzing how attackers gather intelligence and build the knowledge required to mount effective attacks, which is examined next.

\subsubsection{Intelligence and Knowledge}
Attackers conduct intelligence gathering through processes that typically progress from black-box reconnaissance to grey-box discovery and, eventually, to white-box collection. During black-box reconnaissance, adversaries collect passive intelligence by analyzing conference papers, technical blogs, and public vulnerability reports to understand deployment patterns and architectural choices \cite{Reconnaissance, Reconnaissance_Conference, Reconnaissance_PublicResearch, Reconnaissance_PrePrint, Reconnaissance_TechnicalBlogs, Reconnaissance_AdversarialReport}. Active network scanning enables the identification of publicly exposed endpoints and configuration weaknesses, while investigation of application stores and victim websites reveals available machine learning APIs \cite{Reconnaissance_ActiveScanning, Reconnaissance_VictimWebsite, Reconnaissance_AppRepo}. In this phase, attackers interact only through public interfaces, inferring model properties from system outputs and timing behaviors, and launching attacks such as model extraction or model stealing \cite{miao2021machine} despite incomplete knowledge.

With partial internal access, the threat escalates to the grey-box phase. Here, attackers leverage compromised credentials or leaked documentation to map the landscape of machine learning artifacts, including software stacks, feature stores, and deployment pipelines \cite{Discovery, Discovery_DiscoverMLArtifacts}. They interrogate model behavior to deduce model family and output space, and tailor their attack strategies accordingly \cite{Discovery_DiscoverMLModelOntology, Discovery_DiscoverMLModelFamily}. In the context of LLMs, grey-box adversaries extract meta-prompts, probe for hallucinations, and analyze model confidence scores to refine their attacks \cite{Discovery_LLMMetaPromptExtraction, Discovery_DiscoverLLMHallucinations, Discovery_DiscoverAIModelOutputs}. Demonstrations by red teams, such as those conducted by MITRE, have illustrated that partial internal knowledge can be sufficient to design effective adversarial inputs targeting model weaknesses \cite{MITRERedTeamPhysicalCountermeasure}. The ability to align exploitation techniques with specific frameworks further increases the reliability and precision of attacks \cite{Reconnaissance}.

In the white-box phase, adversaries obtain comprehensive access, often through insider roles or compromised administrative credentials \cite{Collection}. Attackers in this position extract models, datasets, and telemetry from internal repositories, thus gaining visibility into data distributions, training procedures, and hyperparameter choices essential for sophisticated and persistent attacks \cite{Collection, Collection_MLArtifactCollection}. Collaborative platforms and local system files provide additional entry points for lateral movement and further data exfiltration \cite{Collection_DatafromInformationRepositories, Collection_DatafromLocalSystem}. White-box access allows for precise computation of adversarial perturbations and targeted data extraction. Incidents such as the Clearview AI breach and disruptive red team exercises targeting cloud machine learning platforms highlight the risks associated with this level of compromise \cite{ClearviewAI, MITRERedTeamMicrosoftAzureDisruption}. The depth of attacker knowledge directly influences both the feasibility and reliability of the techniques that can be deployed \cite{Reconnaissance, Discovery, Collection}. Once attackers acquire the necessary intelligence, the next critical step is to translate this knowledge and motivation into operational capabilities, as discussed below.

\subsubsection{Capabilities and Staging} \label{sec:capabilities_and_staging}
Adversaries develop attack capabilities by acquiring technical knowledge from academic literature and online resources, and by utilizing open-source frameworks such as CleverHans, ART, and Foolbox, which automate adversarial example generation and lower the barrier for executing advanced attacks \cite{cleverhans, art, foolbox}. These frameworks are often combined with general-purpose scripting tools, established machine learning libraries such as PyTorch and TensorFlow, and command-line interpreters that support the automation of exploitation activities \cite{ResourceDevelopment_ObtainCapabilities, CommandandScriptingInterpreter, python}.

To build attack infrastructure, adversaries rent servers, domains, and scalable cloud resources for staging and execution, often selecting GPU-enabled environments provided by Google Colab, AWS, and Azure to support computationally intensive workloads \cite{ResourceDevelopment_AcquireInfrastructure, google_colab, aws, azure, ResourceDevelopment_AcquireInfrastructureMLDevelopmentWorkspaces}. Activities are distributed across multiple accounts and domains to obscure attack traffic and hinder detection. In some cases, attackers exploit virtualization and cloud management vulnerabilities to escalate privileges or pivot to additional targets within the environment \cite{owasp}.

The collection of machine learning artifacts from public repositories represents another critical capability. Adversaries identify datasets and pre-trained models that closely resemble the target environment, using these proxies to develop transferable attack techniques \cite{ResourceDevelopment_AcquirePublicMLArtifact, ResourceDevelopment_AcquirePublicMLArtifactDatasets, ResourceDevelopment_AcquirePublicMLArtifactModels}. The establishment of seemingly legitimate accounts provides a cover for extended reconnaissance and the subsequent staging of attacks \cite{ResourceDevelopment_EstablishAccounts}.

Supply chain compromise is an increasingly prevalent vector for persistent access. Attackers inject malicious packages or dependencies into the software ecosystem, leveraging techniques such as dependency confusion to insert backdoored components through naming collisions in package management systems. The PyTorch dependency confusion incident, which resulted in exfiltration of developer data, exemplifies this threat \cite{MLSupplyChainCompromiseMLSoftware, ResourceDevelopment_ObtainCapabilitiesSoftwareTools, PyTorch_dependency_compromise}. Advanced adversaries may also craft custom attack tools, including adversarial websites, notebook templates embedded with exfiltration mechanisms, or scripts designed to exploit collaborative platforms. Scripting languages such as Python, PowerShell, and Unix shells are common vectors in these workflows \cite{Jupyter, CommandandScriptingInterpreter, python, powershell, owasp}.

The final operational phase centers on staging and delivering the attack. Adversaries publish poisoned datasets and models in public or private repositories, or manipulate training data within established pipelines to introduce backdoor triggers and vulnerabilities \cite{ResourceDevelopment_PublishPoisonedDatasets, ResourceDevelopment_PoisonTrainingData}. Compromised models are uploaded to public registries in the expectation that downstream consumers will adopt them, often during fine-tuning or deployment processes \cite{ResourceDevelopment_PublishPoisonedModels, AttackStaging_BackdoorMLModel}. In LLM-focused environments, attackers fabricate hallucinated entities, such as fictitious packages or domains, to mislead users and security teams. Infrastructure is established with persistent accounts and domains, ensuring continued access even as legitimate users update or replace poisoned artifacts \cite{ResourceDevelopment_AcquireInfrastructureDomains, Persistence, ResourceDevelopment_PoisonTrainingData, AttackStaging_BackdoorMLModel}. Taken together, these capabilities enable attackers to compromise MLOps environments by progressing from motivation and intelligence gathering to the acquisition and staging of operational tools. The following section examines concrete attacks on MLOps and LLMOps systems, and presents defenses and mitigations organized according to these attacker profiles and lifecycle stages.

\begin{figure*}[htbp]
    \includegraphics[scale=0.425]{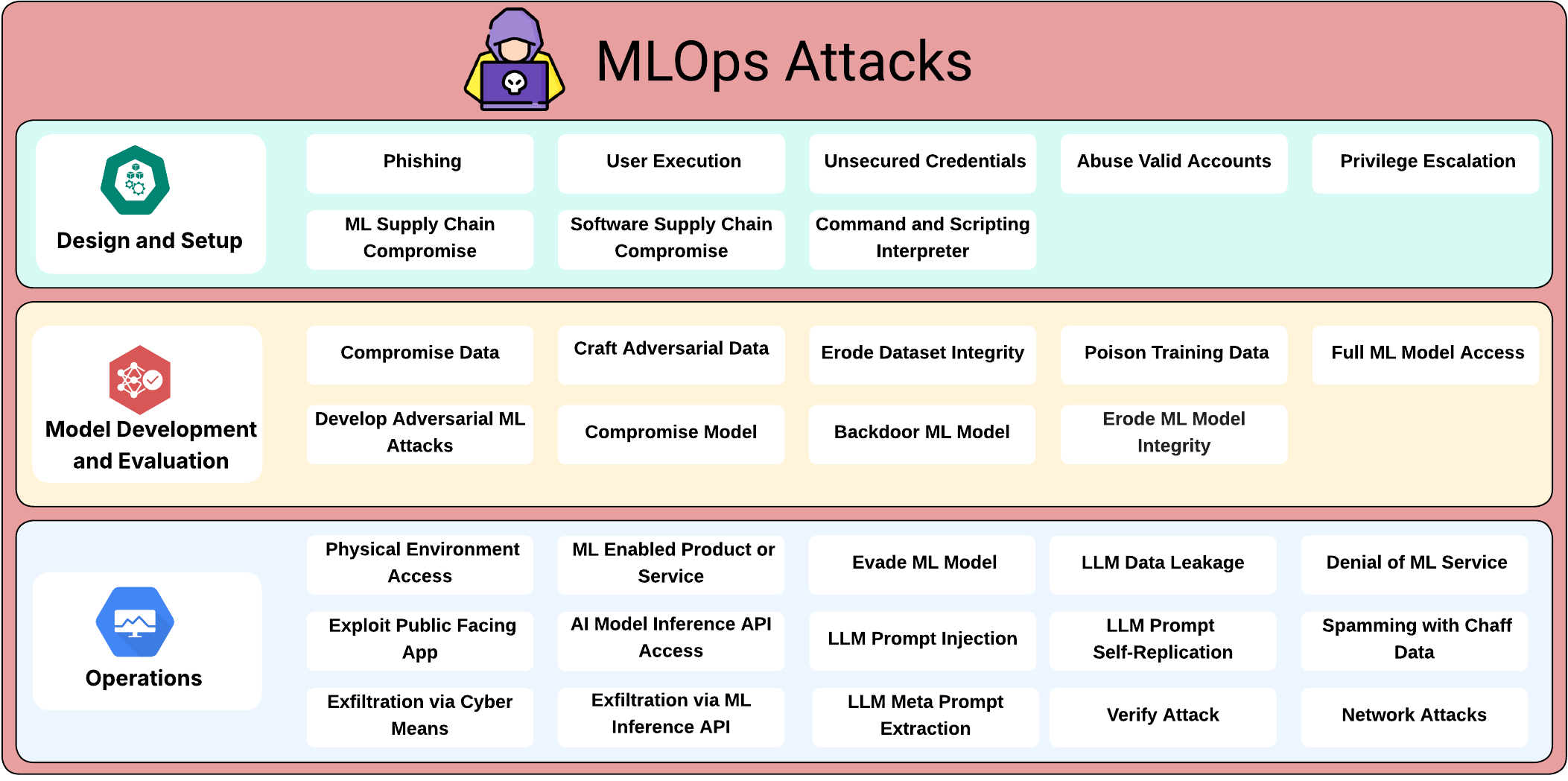}
    \caption{Attacks are organized according to the taxonomic categories presented in Section \ref{sec:survey_taxonomy}. The three groups act as analytical lenses rather than literal or compressed lifecycle phases and support a structured comparison of how diverse threats concentrate around design choices, model-centric workflows, and operational deployment contexts.}
    \Description{Chart categorizing attacks into taxonomical categories previously presented across MLOps pipeline.}
    \label{fig:mitre-attack}
\end{figure*}

\section{Attack and Challenges Review} \label{sec:attack-taxonomy}
Building on the survey approach and threat model described earlier, this section turns to concrete attacks that target the MLOps and LLMOps ecosystems in order to address RQ2. Our analysis is organized around three taxonomical categories introduced in Section~\ref{sec:survey_taxonomy}: \textit{Design and Setup}, \textit{Model Development and Evaluation}, and \textit{Operations}, which are depicted in Fig.~\ref{fig:mitre-attack}. For each category, we map relevant attack techniques from the MITRE ATLAS framework to real-world incidents and red-teaming exercises, distinguishing between threats that have materialized in practice and those that remain theoretical. This evidence-based approach aims to give practitioners actionable intelligence about the threat landscape and inform both organizational and technical risk management decisions. Across all three categories, attack paths frequently manifest as supply chain attacks that target users or systems, which highlights the interconnected and layered nature of vulnerabilities present throughout the MLOps lifecycle.

\subsection{Attacks in Design and Setup}
The \textit{Design and Setup} phase is where organizational and technical choices establish the foundation for the MLOps environment. Attacks at this stage often exploit weaknesses in human factors, identity management, configuration, and core infrastructure to gain initial access to machine learning systems. Human users are frequently regarded as the weakest link in security \cite{human_weakest_link}, and attackers exploit this reality by combining social engineering with technical manipulation. Their goal is to escalate from basic access to more advanced forms of exploitation. Initial access is commonly gained through phishing, social engineering, user execution, and privilege escalation, prompting users to take unintentional actions that activate malicious code or introduce unsafe packages, which ultimately undermines operational integrity.

Phishing remains one of the most effective entry points into MLOps environments \cite{phishing, Phishing_mitre}. Traditional phishing relies on generic fraudulent communications, while spearphishing targets individuals with highly personalized messages \cite{Phishing_SpearphishingviaSocialEngineeringLLM}. The evolution of these attacks is evident in recent incidents where generative AI and deepfake technologies are used to craft sophisticated, context-aware messages that closely mimic legitimate organizational communication \cite{ai_phishing, Phishing_SpearphishingviaSocialEngineeringLLM}. The Computronix AI Phishing Campaign highlights this shift. In this incident, attackers leveraged automation and natural language generation to create highly personalized messages that bypass spam filters and deceive even vigilant, security-conscious users \cite{ai_phishing}. These AI-powered campaigns allow attackers to scale their operations efficiently, preserving context-specific targeting and achieving higher success rates than traditional phishing attempts. In parallel, user execution attacks \cite{Execution_UserExecution} manipulate users into executing malicious code, often introduced through compromised components in the machine learning supply chain \cite{MLSupplyChainCompromise}. Unsafe ML artifacts can carry harmful payloads, while serialization vulnerabilities in model storage or loading routines may be exploited for code execution if proper validation is lacking \cite{Execution_UserExecution_UnsafeMLArtifacts, owasp}.

An illustrative case is the PyTorch Dependency Confusion attack, where attackers uploaded a malicious package to the public PyPI repository. This package impersonated a legitimate PyTorch dependency and, during routine updates, was automatically installed by many users. As a result, sensitive system information and SSH keys were exfiltrated to attacker-controlled servers \cite{Execution_UserExecution, PyTorch_dependency_compromise, Execution_UserExecution_MaliciousPackage}. The incident highlights how trust in software package ecosystems can be weaponized through what appear to be routine update procedures. Similarly, the ChatGPT Package Hallucination incident exploited hallucinations in LLM outputs by registering non-existent packages that users, relying on LLM recommendations, subsequently installed. This led to compromise of their systems \cite{ExfiltrationviaMLInferenceAPI_InferTrainingDataMembership}.

Privilege escalation attacks represent another powerful tool for attackers, enabling them to expand access and achieve broader objectives within the MLOps environment \cite{PrivilegeEscalation}. Such escalation may result from LLM prompt injection \cite{LLMPromptInjectionDirect, LLMPromptInjectionInDirect}, LLM plugin compromise \cite{LLMPluginCompromise}, or jailbreak techniques \cite{LLMJailbreak}. The ChatGPT Plugin Privacy incident demonstrated how excessive plugin permissions can become a critical vulnerability. In this scenario, attackers used markdown injection to embed malicious links that exfiltrated sensitive conversation data from users \cite{PrivilegeEscalation, chatgpt_privacy_leak}. The Morris II worm highlighted the risk of indirect prompt injection, where retrieval-augmented generation in an email assistant enabled self-replication and automated exfiltration of sensitive data across multiple systems \cite{Morris_II}. These examples demonstrate the necessity for strong privilege boundaries and rigorous access controls.

Attacks targeting access abuse naturally follow, as attackers take advantage of legitimate credentials or authentication tokens. These include API keys, service tokens, and cloud account credentials, which allow them to bypass conventional security controls and manipulate machine learning infrastructure \cite{CredentialAccess_mitre}. Weakly managed credentials such as default accounts, dormant accounts, and cloud credentials present attractive opportunities \cite{ValidAccounts, ValidAccounts_mitre, ValidAccounts_DefaultAccounts_mitre, ValidAccounts_LocalAccounts_mitre, ValidAccounts_CloudAccounts_mitre}. During a red-teaming exercise on Microsoft Azure ML services, attackers used compromised credentials to escalate from basic user privileges to administrative access. They ultimately corrupted model outputs and exfiltrated sensitive training data \cite{MITRERedTeamMicrosoftAzureDisruption}. Similarly, a Google Colab vulnerability enabled arbitrary code execution, which allowed attackers to infiltrate internal services and escalate privileges through exposed cloud credentials \cite{google_colab_code_execution}.

Unsecured credential exposure further amplifies these risks, especially when secrets are stored in plaintext, in source code, or as environment variables across cloud and containerized environments \cite{UnsecuredCredentials, UnsecuredCredentials_mitre}. The MathGPT incident illustrates this challenge: exposed API keys and authentication tokens enabled attackers to access high-value resources and sensitive training data, causing significant financial and security damage \cite{mathgpt_compromise}. Attackers also harvest secrets by exploiting access to cloud instance metadata APIs, Kubernetes clusters, Docker images, and even collaboration tools such as Slack and Trello \cite{UnsecuredCredential_CloudMetadataAPI_mitre, kubernetes, docker, slack, Trello}. Automated tools and scripts have made credential harvesting highly efficient, and the Ray framework vulnerability showed how widespread the consequences can be when thousands of deployments are exposed at scale \cite{ShadowRay}. The difficulty in distinguishing malicious from legitimate use is exacerbated when credential hygiene is poor and monitoring is insufficient.

Beyond these risks, infrastructure attacks focus on the critical systems that support machine learning workflows. These include servers, CI/CD pipelines, GPUs, version control systems, and collaborative environments \cite{7445136}. Compromising any of these components allows attackers to manipulate or delay model training, testing, or deployment. Many attacks involve the insertion of malicious code through scripting languages commonly used in MLOps, such as Python and PowerShell \cite{CommandandScriptingInterpreter, python, powershell}. Because these scripts underpin core CI/CD workflows, attackers can disguise their activities within routine operations, making detection challenging in environments that depend on continuous integration and deployment \cite{owasp}. Hardware compromise and malicious software supply chain attacks can introduce adversarial examples or backdoored libraries that persist throughout the ML lifecycle \cite{MLSupplyChainCompromiseHardware, MLSupplyChainCompromiseMLSoftware, ResourceDevelopment_ObtainCapabilitiesSoftwareTools}. Together, these attacks reveal how initial setup and configuration decisions can propagate vulnerabilities that undermine the integrity of machine learning operations. The next subsection examines attacks that target the core processes of model development and evaluation.

\subsection{Attacks in Model Development and Evaluation}
The \textit{Model Development and Evaluation} phase faces attacks that compromise data integrity, model behavior, and the security of training processes. Unlike threats in the \textit{Design and Setup} phase that primarily aim to obtain access, adversarial techniques here focus on corrupting the machine learning artifacts themselves. These artifacts include training data, fine-tuning data, labels, model architectures, weight parameters, and pre-trained components. Such attacks introduce subtle and persistent vulnerabilities that may remain undetected during evaluation and continue to influence system behavior throughout deployment and production use. Because modern ML development depends on large-scale automated data pipelines, public datasets, and the reuse of shared model components, this phase becomes a critical point where attackers can shape system behavior at scale and with long-lasting effect.

A central adversarial objective is the degradation of data integrity, since both training and operational data must remain accurate, representative, and trustworthy. Poisoning attacks embed harmful samples or misleading content into pre-training or fine-tuning datasets, thereby degrading model performance or embedding hidden behaviors that activate under specific conditions \cite{ResourceDevelopment_PoisonTrainingData, owasp}. Early incidents such as the Microsoft Tay Chatbot failure demonstrated the risks of unfiltered continuous learning, where coordinated adversaries exploited the system’s exposure to public social media interactions and triggered rapid degradation of model behavior \cite{tay_poisoning}. More advanced poisoning strategies such as split-view poisoning exploit predictable data collection schedules at web scale. Attackers inject malicious content into webpages shortly before automated crawlers run, ensuring poisoned data enters the training pipeline even though ordinary users observe a different version of the webpage \cite{Poisoning_Web-Data}. Experiments have shown that poisoning as little as 0.01 percent of a dataset can produce targeted misclassifications while headline accuracy metrics remain stable, making detection through standard validation extremely difficult \cite{Impact_ErodeDatasetIntegrity}.

Beyond direct poisoning, training-only manipulation attacks such as feature collision, label flipping, and bilevel optimization introduce subtle vulnerabilities that bypass conventional detection \cite{dataset_security}. Federated learning systems amplify these risks because poisoned weights can be injected into the aggregation process from multiple decentralized nodes, complicating attribution and undermining global model robustness. Attackers may also insert false labels, malware samples, or mislabeled content into open-source datasets, causing organizations to unknowingly import tainted samples into their pipelines \cite{ResourceDevelopment_PublishPoisonedDatasets, yerlikaya2022data}. These threats extend into the inference stage, where adversarial inputs crafted under black-box and white-box conditions induce misclassification without modifying training data. Embedded backdoor triggers enable additional targeted manipulation by activating hidden behaviors only when specific patterns or tokens appear in inputs \cite{CraftAdversarialData_InsertBackdoorTrigger}.

Real-world incidents illustrate how such adversarial techniques accumulate over time to degrade entire security ecosystems. The VirusTotal poisoning campaign revealed how systematically submitting borderline malware samples could shift decision boundaries for multiple commercial detection models, eventually enabling new malware variants to evade detection across the industry \cite{VirusTotalPoisoning}. Similarly, adversarial capabilities demonstrated by Kaspersky researchers showed how perturbations crafted against local feature extractors could transfer successfully to cloud-based antimalware systems, proving that attackers can develop and test attacks locally before deploying them at scale against remote services \cite{kaspersky_adversarial_attack}. These insights demonstrate that adversaries frequently refine their attack strategies through the use of open-source adversarial ML frameworks. Toolkits such as CleverHans, ART, and Foolbox automate adversarial example generation, backdoor construction, and robustness evaluation, lowering the barrier for developing advanced attacks and enabling even relatively low-resourced adversaries to reproduce sophisticated techniques \cite{cleverhans, art, foolbox}.

While data integrity attacks focus on corrupting training inputs, model integrity attacks target the internal architecture, learned representations, and distribution mechanisms of ML models. These attacks degrade decision-making capability, create opportunities for the extraction of private information, and enable the covert spread of misinformation. Backdoor attacks embed malicious triggers into models through poisoned data or direct parameter manipulation, causing the system to behave normally under standard evaluation yet reveal attacker-controlled behavior when specific inputs appear \cite{AttackStaging_BackdoorMLModel}. Hidden triggers undermine reliability and force organizations to expend significant resources investigating and repairing corrupted models \cite{Impact_ErodeMLModelIntegrity}. When adversaries possess full access to model architectures and parameters, they can design highly precise backdoors or payloads customized for particular tasks, users, or operational environments \cite{FullMLModelAccess}.

Deep payload injection represents one of the most concerning manifestations of model integrity compromise. In this attack, executable code is embedded directly into neural network weight matrices, converting a seemingly benign ML model into a covert malware delivery mechanism \cite{deeppayload}. Researchers identified dozens of vulnerable Android applications whose on-device models performed tasks such as cash recognition, user authentication, parental control, and financial processing. Although the corrupted models produced normal inference outputs, they enabled data exfiltration, command execution, and unauthorized behavioral modification. Detection through functional testing became practically impossible because the underlying prediction behavior remained intact.

Model supply chain compromise magnifies these risks by targeting pre-trained and publicly shared models. Threat actors have uploaded malicious or backdoored models to widely used repositories such as Hugging Face, where unsuspecting developers may integrate them directly into downstream systems without realizing that dormant triggers or embedded payloads remain hidden within the model \cite{ResourceDevelopment_PublishPoisonedModels}. The PoisonGPT demonstration exemplified this risk by releasing a modified large language model that passed standard benchmark evaluations but returned targeted misinformation when queried about specific topics \cite{PoisonGPT}. Because such attacks exploit the trust placed in public model hubs and the widespread reuse of pre-trained model components, compromised artifacts can propagate across organizations and industries. Outdated or poorly maintained models further increase exposure because they often lack important security updates and remain vulnerable to parameter editing techniques, including emerging methods like Rank-One Model Editing (ROME) \cite{owasp}.

Taken together, these attacks show how adversaries can quietly and systematically undermine both the data foundations and the computational logic of ML systems during development and evaluation. By corrupting datasets, manipulating learned representations, and compromising pre-trained artifacts distributed across shared ecosystems, attackers influence downstream predictions long after deployment. The cumulative effect is an erosion of trust in model reliability, increased vulnerability to operational exploitation, and long-term degradation of system security. Having examined these foundational weaknesses in model development and evaluation, the next subsection turns to attacks in the Operations phase, where adversaries exploit compromised models and the surrounding infrastructure during deployment and real-time interaction.

\subsection{Attacks in Operations}
The \textit{Operations} phase exposes deployed machine learning and large language model systems to direct attacker interaction through public APIs, user interfaces, and network endpoints. In this stage, attackers leverage runtime vulnerabilities to extract information, evade detection, or disrupt mission-critical services. Operational attacks often combine techniques across physical, application, and infrastructure layers, with these actions unfolding in real time and leading to significant consequences. As machine learning and large language model deployments continue to grow in various domains, the scale and severity of operational risks have become more pronounced, making it critical to examine how these attacks manifest within modern system architectures \cite{Discovery_DiscoverAIModelOutputs, AIModelInferenceAPIAccess}.

Physical manipulation of sensor inputs represents a significant operational risk, especially for systems that process real-world data. Bypassing biometric authentication is possible when physical security is circumvented using engineered artifacts. For example, red team researchers have achieved targeted misidentification in facial recognition systems using printed masks and infrared LEDs, resulting in both impersonation, where unauthorized users gain access, and anonymization, where legitimate users are not recognized \cite{PhysicalEnvironmentAccess, MITRERedTeamPhysicalCountermeasure}. These events demonstrate that attackers can compromise machine learning systems using physical means alone, without the need for digital intrusion. Identity verification platforms are also susceptible to attacks that exploit synthetic media such as deepfakes. In the Bypassing ID.me incident, adversaries were able to defeat liveness detection and identity matching mechanisms, which allowed them to fraudulently claim government benefits and convert these weaknesses into financial gain \cite{BypassID.ME, ML-EnabledProductorService}.

Attacks on service-level and infrastructure components target the core platforms and tools that support machine learning workloads. Publicly accessible APIs, as shown by the ShadowRay exploitation campaign, provide attackers with opportunities to exploit unauthenticated endpoints in cloud-based deployments. Adversaries in this case executed arbitrary code, deployed cryptocurrency miners, exfiltrated sensitive training data, and established persistent access by targeting Ray Jobs APIs, which led to the compromise of more than five thousand servers and significant financial and operational losses \cite{ShadowRay, ExploitPublic-FacingApplication}. These attacks are not limited to resource theft. Attackers have also initiated resource exhaustion campaigns, as demonstrated in the MathGPT incident, where a lack of rate limiting allowed automated large-scale API queries to overwhelm computational resources and increase operational costs \cite{Impact_DenialofMLService, mathgpt_compromise}. Collaborative environments such as Google Colab are at risk for code execution attacks, in which adversaries inject malicious scripts into shared notebooks. When these notebooks are opened, attackers can exfiltrate datasets, credentials, and proprietary code, resulting in the compromise of research efforts \cite{ExfiltrationviaCyberMeans, google_colab_code_execution}.

A critical area of operational attacks centers on the confidentiality and privacy of training data, model internals, and user information. Attackers have developed techniques to infer whether certain data samples were included in training sets and to extract global statistical properties or reconstruct portions of the original data. These methods often use shadow models or detailed analysis of prediction scores \cite{ExfiltrationviaMLInferenceAPI_InferTrainingDataMembership}. Such data exfiltration through inference APIs introduces substantial privacy concerns, as models may reveal personally identifiable information or intellectual property present in training data. Large language models are particularly vulnerable to data leakage attacks in which specific prompts elicit sensitive or proprietary information, including material from proprietary corpora, public sources, or even previous user sessions. These vulnerabilities have become major security concerns for interactive machine learning and large language model deployments, since models with high capacity may unintentionally reveal information that affects organizational trust and regulatory standing.

Operational risks also include model extraction attacks, where adversaries systematically query inference APIs to replicate proprietary models or steal intellectual property. In the Machine Learning Translation attack, researchers used strategically designed API queries to clone commercial translation systems, effectively recreating models that required years of investment \cite{google_translation, AIModelInferenceAPIAccess}. Evasion attacks further challenge operational security by introducing adversarial inputs that can bypass detection. Studies of domain generation algorithm detectors and AI-powered antivirus tools such as Cylance have shown that small character-level or byte sequence modifications can defeat these models while maintaining malicious intent \cite{VerifyAttack, DGA_detection, Impact_EvadeMLModel, cylance2019}. Attackers typically validate such methods in controlled or low-risk environments before targeting production systems, allowing them to refine attack strategies and reduce the chance of early discovery \cite{VerifyAttack}.

Large language models expand the operational attack surface because they rely on natural language interfaces and often integrate with third-party tools. Prompt injection attacks use conversational context or document uploads to insert hidden instructions, which may cause models to leak sensitive information, produce misleading outputs, or circumvent safety measures \cite{LLMPromptInjection, AgentFlayer2025}. The Morris II worm incident is an example of a self-replicating attack in large language model applications. Here, malicious prompts autonomously spread by extracting and forwarding context to new recipients, enabling the infection to propagate across different organizations \cite{LLMDataLeakage, LLMPromptSelf-Replication, Morris_II}. Meta prompt extraction attacks allow adversaries to discover hidden system instructions or internal logic, which can expose proprietary knowledge or change how the model behaves \cite{Discovery_LLMMetaPromptExtraction, owasp}. Incidents such as these demonstrate that products and services powered by machine learning are frequent targets, and attackers continue to search for operational entry points and develop attack methods specific to the machine learning lifecycle \cite{ML-EnabledProductorService}.

Network exploitation in the operational phase is another critical threat, where attackers target interconnected infrastructure to disrupt services, compromise data, or degrade pipelines \cite{Impact_DenialofMLService, chaff-original}. By taking advantage of exposed endpoints and weak communication protocols, attackers can manipulate workflows, exfiltrate information, inject poisoned or malformed data, or make services unavailable for legitimate users \cite{ExfiltrationviaCyberMeans, ExploitPublic-FacingApplication, chaff-original}. Notable incidents involving network-based attacks, including those targeting Ray Jobs APIs or overloading systems with chaff data, reveal how vulnerabilities in a single area can cascade and put entire platforms at risk. Attackers may also gain unauthorized access by hijacking sessions, forging credentials, or manipulating network configurations. In distributed machine learning environments, proactive measures such as robust network segmentation, ongoing traffic monitoring, anomaly detection, and strong authentication protocols are needed to mitigate these risks \cite{WebProtocols_mitre, ReflectionAmplification_mitre, SessionHijacking_mitre, ForgeWebCredentials_mitre, DHCPSpoofing_mitre}.

Taken together, these operational attacks show that machine learning and large language model systems are exposed to continuous and evolving threats after deployment. The scope of attacks includes physical sensor manipulation, API abuse, code execution, data and model exfiltration, adversarial querying, prompt injection, meta prompt extraction, and complex network exploitation. When analyzed alongside attacks in the Design and Setup as well as Model Development and Evaluation phases, which focus on human factors, data pipelines, and model artifacts, the operational phase reveals the entire spectrum of layered risks present in today’s MLOps and LLMOps environments. Attackers rely on both conventional information technology techniques, such as phishing and credential theft, and machine learning-specific strategies, including poisoning, backdoors, model extraction, adversarial querying, and prompt injection. Real-world incidents and red teaming exercises have demonstrated that these threats are not only theoretical but have resulted in measurable operational, financial, and reputational damage. Table~\ref{tab:attack-validation} presents a synthesis of attack techniques observed across the machine learning lifecycle, their empirical validation, and how they overlap with traditional information technology security concerns. The following section transitions from the analysis of attacks to an in-depth discussion of defenses and mitigation strategies designed to strengthen MLOps and LLMOps deployments against persistent adversarial threats.

\begin{table*}[htbp]
    \caption{Summary of MLOps attack techniques with empirical validation. Traditional IT column indicates attacks that overlap with conventional cybersecurity (\ding{51}) or are unique to ML environments (\ding{53}). The empirical validation column cites real world incidents or red team exercises that demonstrate practical exploitation of each technique.}
    \begin{adjustbox}{max width=\textwidth}
        \begin{tabular}{|p{2cm}|p{5cm}|p{4.8cm}|p{2cm}|}
\hline
\rowcolor[HTML]{E1FFFF} 
{\color[HTML]{010101} \textbf{Taxonomy}} & {\color[HTML]{010101} \textbf{Attack Techniques}} & {\color[HTML]{010101} \textbf{Empirical Validation}} & {\color[HTML]{010101} \textbf{Traditional IT}} \\ \hline

& Phishing \cite{phishing} & AI Phishing Campaign \cite{ai_phishing} & \ding{51} \\ \cline{2-4} 
& Unsecured Credentials \cite{UnsecuredCredentials} & MathGPT incident \cite{mathgpt_compromise} & \ding{51} \\ \cline{2-4} 
& Privilege Escalation \cite{PrivilegeEscalation} & ChatGPT Plugin Privacy \cite{chatgpt_privacy_leak} & \ding{51} \\ \cline{2-4} 
& User Execution \cite{Execution_UserExecution} & PyTorch Dependency \cite{PyTorch_dependency_compromise} & \ding{51} \\ \cline{2-4} 
& Abuse Valid Accounts \cite{ValidAccounts} & Azure Service Disruption \cite{MITRERedTeamMicrosoftAzureDisruption} & \ding{51} \\ \cline{2-4} 
\multirow{-6}{2cm}{Design and Setup} & ML Supply Chain Compromise & HuggingFace Malicious Model \cite{MaliciousMLModel_HuggingFace} & \ding{51} \\ \hline

& Compromise Data \cite{MLSupplyChainCompromiseData} & Microsoft Tay Chatbot \cite{tay_poisoning} & \ding{53} \\ \cline{2-4} 
& Erode Dataset Integrity \cite{Impact_ErodeDatasetIntegrity} & Split view poisoning \cite{Poisoning_Web-Data} & \ding{53} \\ \cline{2-4} 
& Poison Training Data \cite{ResourceDevelopment_PoisonTrainingData} & Split view poisoning \cite{Poisoning_Web-Data} & \ding{53} \\ \cline{2-4} 
& Compromise Model \cite{MLSupplyChainCompromiseModel} & Deep Payload Injection \cite{deeppayload} & \ding{53} \\ \cline{2-4} 
& Develop Adversarial ML Attacks \cite{ResourceDevelopment_DevelopCapabilitiesAdversarialMLAttacks} & Kaspersky Adversarial Attack \cite{kaspersky_adversarial_attack} & \ding{53} \\ \cline{2-4} 
& Craft Adversarial Data \cite{CraftAdversarialData} & VirusTotal poisoning \cite{VirusTotalPoisoning} & \ding{53} \\ \cline{2-4} 
& Backdoor ML Model \cite{AttackStaging_BackdoorMLModel} & PoisonGPT \cite{PoisonGPT} & \ding{53} \\ \cline{2-4} 
& Erode ML Model Integrity \cite{Impact_ErodeMLModelIntegrity} & VirusTotal poisoning \cite{VirusTotalPoisoning} & \ding{53} \\ \cline{2-4} 
\multirow{-9}{2cm}{Model Development and Evaluation} & Full Model Access & Deep Payload Injection \cite{deeppayload} & \ding{53} \\ \hline

& Physical Environment Access \cite{PhysicalEnvironmentAccess} & Face Identification System \cite{MITRERedTeamPhysicalCountermeasure} & \ding{53} \\ \cline{2-4} 
& Exploit Public Facing Application \cite{ExploitPublic-FacingApplication} & ShadowRay Exploitation \cite{ShadowRay} & \ding{51} \\ \cline{2-4} 
& ML Enabled Product or Service \cite{ML-EnabledProductorService} & Bypassing ID.me \cite{BypassID.ME} & \ding{53} \\ \cline{2-4} 
& AI Model Inference API Access \cite{AIModelInferenceAPIAccess} & Machine Learning Translation \cite{google_translation} & \ding{53} \\ \cline{2-4} 
& Verify Attack \cite{VerifyAttack} & DGA Detection Evasion \cite{DGA_detection} & \ding{53} \\ \cline{2-4} 
& Spamming with Chaff Data \cite{Impact_SpammingMLSystemwithChaffData} & \ding{53} & \ding{53} \\ \cline{2-4} 
& Denial of ML Service \cite{Impact_DenialofMLService} & MathGPT incident \cite{mathgpt_compromise} & \ding{53} \\ \cline{2-4} 
& Exfiltration via Cyber Means \cite{ExfiltrationviaCyberMeans} & Google Colab Code Execution \cite{google_colab_code_execution} & \ding{51} \\ \cline{2-4} 
& Exfiltration via ML Inference API \cite{ExfiltrationviaMLInferenceAPI} & \ding{53} & \ding{53} \\ \cline{2-4} 
& Evade ML Model \cite{Impact_EvadeMLModel} & Cyclance AI Malware Detection \cite{cylance2019} & \ding{53} \\ \cline{2-4} 
& LLM Prompt Injection \cite{LLMPromptInjection} & AgentFlyer Copilot \cite{AgentFlayer2025} & \ding{53} \\ \cline{2-4} 
& LLM Meta Prompt Extraction \cite{Discovery_LLMMetaPromptExtraction} & \ding{53} & \ding{53} \\ \cline{2-4} 
& LLM Data Leakage \cite{LLMDataLeakage} & Morris II worm \cite{Morris_II} & \ding{53} \\ \cline{2-4} 
\multirow{-14}{2cm}{Operations} & LLM Prompt Self-Replication & Morris II worm \cite{Morris_II} & \ding{53} \\ \hline
\end{tabular} \label{tab:attack-validation}
    \end{adjustbox}
    \label{tab:attacks-mapping}
\end{table*}

\begin{figure*}[htbp]
    \includegraphics[scale=0.425]{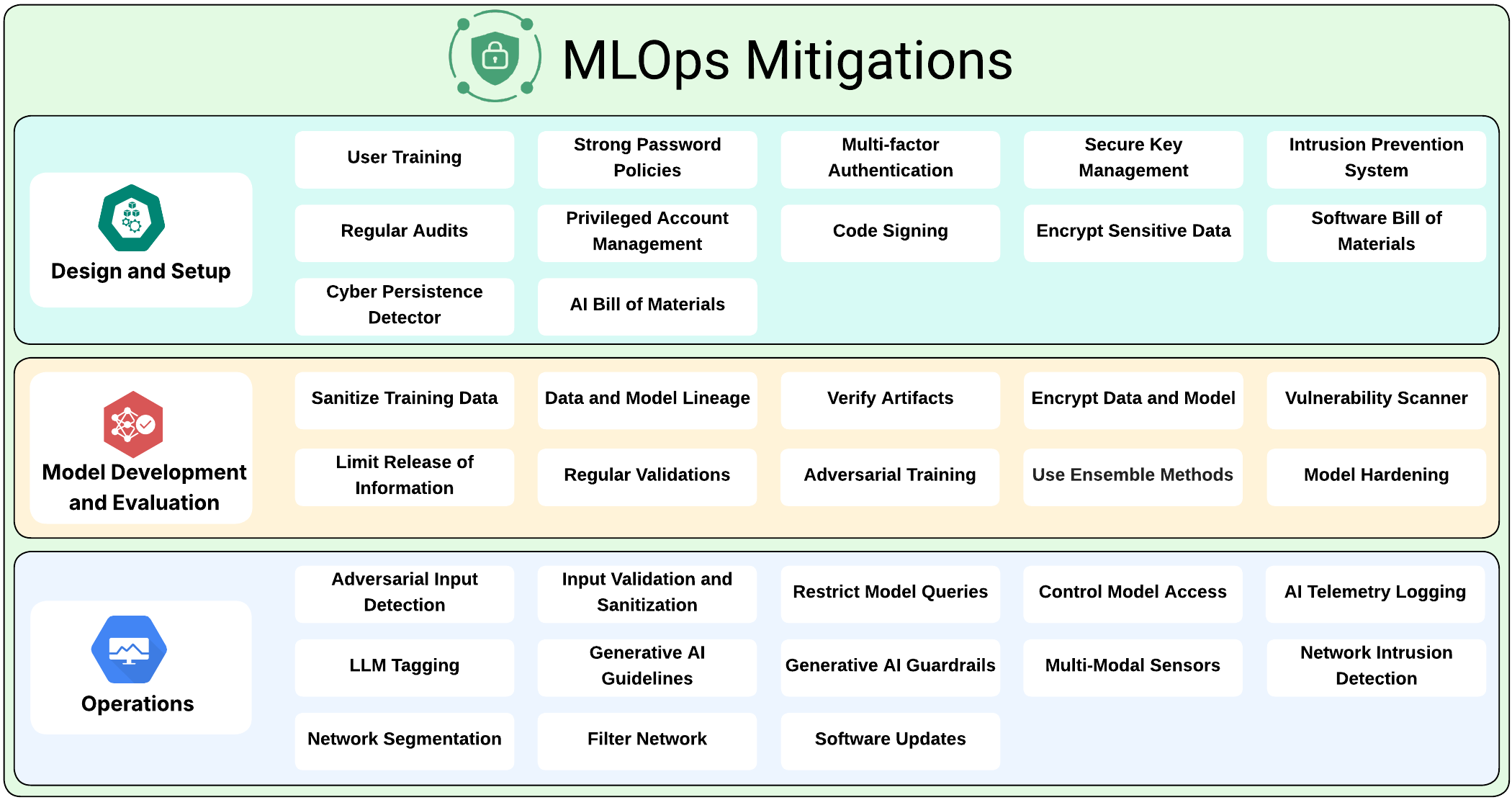}
    \caption{Mitigation strategies organized according to the taxonomic categories presented in Section \ref{sec:survey_taxonomy}, corresponding to the attacks shown previously. These categories enable systematic alignment of defenses with design choices, model-centric workflows, and operational deployment contexts.}
    \Description{Chart categorizing defenses into the same taxonomical categories previously presented.}
    \label{fig:mitre-defenses}
\end{figure*}

\section{Security Mitigation Strategies} \label{sec:mitigations-taxonomy}
The previous section outlined the attack landscape and identified key vectors that attackers may exploit across the MLOps lifecycle. This section turns to corresponding mitigation strategies. To address RQ3, we assess how well the mitigation techniques proposed in existing frameworks align with the identified attack techniques and whether they provide comprehensive coverage of the resulting attack surface. Drawing from MITRE ATLAS, MITRE ATT\&CK, and recent research, we group defenses by the same three categories used in the attack taxonomy. Fig.~\ref{fig:mitre-defenses} summarizes the mitigation techniques associated with each phase of the lifecycle and serves as a visual reference for the discussion that follows. For each phase we highlight concrete mitigation techniques, relate them to real world deployments, and explain how they jointly constrain attacker opportunities introduced in Section~\ref{sec:attack-taxonomy}. We begin by examining mitigations for the Design and Setup phase, which establish the security foundation for the entire MLOps pipeline.

\subsection{Mitigations for Design and Setup} \label{sec:design_and_setup_mitigations}
Mitigations in the \textit{Design and Setup} phase focus on preventing initial compromise and limiting the impact of successful entry into the environment. As indicated in Fig.~\ref{fig:mitre-defenses}, these defenses span user-focused training, administrative controls, and infrastructure-level safeguards that collectively shape the security posture of the MLOps and LLMOps stack. The remainder of this subsection explains how these classes of mitigation reduce opportunities available to attackers and how they have been applied in practice, progressing from user-facing controls through identity management to technical infrastructure protections and supply chain security.

User-facing mitigations address phishing, social engineering, and unsafe execution behavior through a combination of education and technical controls that work together to create multiple barriers against initial compromise. User training \cite{UserTraining} forms the foundation of human-centered defenses, with security awareness training programs recommended to provide at least quarterly training sessions and simulated phishing tests to reduce organizational risk \cite{SecurityAwarenessTraining}. Organizations can further reduce phishing exposure and malicious code execution by deploying antivirus and antimalware tools, running regular audits and scans \cite{audit_mitre}, and blocking suspicious traffic with intrusion prevention systems \cite{BehaviorPreventiononEndpoint_mitre}. A study of 100 companies in Indonesia that implemented IT audits reported that 85\% of organizations saw reduced risk of data breaches and 78\% observed improved operational efficiency \cite{inplimentationofinformation}. A classic Cisco case study shows how network-based intrusion prevention detected and mitigated internal security events before users experienced secondary impact \cite{Reid1992}. These technical controls complement user education by providing automated defenses that can detect and block threats even when human vigilance fails, creating a defense-in-depth approach where multiple layers of protection work in concert to reduce the organization's attack surface.

Beyond user education and automated threat detection, strong identity and account controls serve as a critical barrier against credential theft and abuse of valid accounts. Strong password policies \cite{PasswordPolicies_mitre} establish baseline credential security, with one case study documenting a move from a six-character default to an eight-character minimum based on empirical work showing that eight-character passwords are much harder to break \cite{globinfo}. Multi-factor authentication \cite{Multi-factorAuthentication_mitre} provides a more robust defense, with a large-scale benchmark of Microsoft Azure Active Directory (AD) users finding that over 99.99\% of multi-factor authentication-enabled accounts remained secure and that this control reduces the risk of compromise by 99.22\% across the population \cite{Meyer2023}. Privileged account management \cite{PrivilegedAccountManagement_mitre} further reduces risk by isolating administrative access from standard user workflows. A practical hardening step is to isolate all privileged accounts from standard user access and mandate multi-factor authentication for privileged access \cite{completeguide}. These identity controls work synergistically, with password policies providing baseline security, multi-factor authentication adding a second verification factor, and privileged account management ensuring that even compromised credentials have limited scope for damage through enforcement of least privilege, separation of duties, and behavioral analytics that help contain abuse when credentials are stolen, including temporal and sequence-aware models for detecting anomalous behavior over time \cite{mehmood2023privilege}.

Building on these identity controls, protecting credentials and configuration data also requires technical safeguards on stored secrets, system configuration, and data at rest and in transit. Encrypting sensitive information \cite{EncryptSensitiveInformation_mitre} reduces disclosure risks, with an enterprise encryption blueprint describing this strategy as protecting data in motion and at rest, with encryption keys as the primary asset to protect \cite{theenterprise}. Secure key management using Hardware Security Modules and Trusted Platform Modules can further protect cryptographic keys \cite{mishra2025modern}, as seen in cloud key management services that support bring your own keys and hold your own keys models to keep customer keys under tighter control \cite{cloudkeymanagement}. These encryption and key management practices must be complemented by access controls and system hardening to prevent unauthorized access to the underlying infrastructure. Restricting resource access over networks, enforcing restrictive operating system configurations, and restricting file and directory permissions narrow the opportunities for lateral movement once an attacker gains initial access. Routine software updates reduce the number of exploitable vulnerabilities, behavior prevention on endpoints blocks malicious execution patterns, execution prevention stops unauthorized code from running, disabling unused features reduces the attack surface, and limiting software installation prevents the introduction of malicious or vulnerable components. Together, these technical safeguards create a hardened environment where credentials and sensitive data are protected through multiple layers of encryption, access control, and system configuration, while continuous monitoring and updating ensure that defenses remain effective against evolving threats.

The integrity of the MLOps environment depends not only on protecting existing infrastructure but also on ensuring that software and AI artifacts entering the system can be trusted, which requires supply chain-focused mitigations that verify the provenance and integrity of all components. A software Bill of Materials \cite{securingthesw} catalogs all software components and dependencies involved in building systems, which enables rapid impact assessment when new vulnerabilities are disclosed. Recent large-scale incidents such as the SolarWinds and Log4j attacks highlight the importance of this visibility into the software supply chain. Code signing \cite{CodeSigning} is used to guarantee that software has not been altered, with a representative deployment centralizing signing profiles so that all code running on a serverless platform such as AWS Lambda must be signed by a trusted developer before execution \cite{bestpracticesandadvanced}. For ML systems specifically, an AI Bill of Materials \cite{AIBillofMaterials} extends this concept to track datasets, models, and training artifacts. In practice, the Neural Network Bill of Materials concept has been applied at scale to 55,997 GitHub repositories to move from conceptual AI Bills of Materials to concrete dependency tracking \cite{Ren2025}. Verifying ML artifacts through cryptographic checksums mitigates risks of poisoned datasets or altered models, with a recent GPU-based framework named Sentry demonstrating practical cryptographic verification for ML artifacts, performing on-the-fly checks and achieving orders of magnitude speedup over CPU-based verification \cite{Gan2025}. These supply chain mitigations provide critical assurance that both traditional software components and ML-specific artifacts can be traced to trusted sources and have not been tampered with during distribution or storage.

Finally, detecting long-lived persistence requires correlating events over time to identify attack patterns that span multiple stages of compromise and remain hidden within normal operational activity. Cyber Persistence Detector (CPD) \cite{liu2024accurate} exemplifies this class of defenses. CPD links setup and execution events with pseudo-dependency edges, uses expert-defined edges and rules to flag suspicious chains of activity, and correlates related kill chain events into concise attack graphs. An alert budget system prioritizes high-value events for analyst review and focuses on behavioral patterns that are harder for attackers to evade than simple indicators. This type of persistence-focused monitoring is particularly important for MLOps environments where model and data assets live across many services and machines, creating numerous opportunities for attackers to establish persistent access that spans compute nodes, storage systems, and orchestration platforms. By correlating events across these distributed components, persistence detection systems can identify attacker footholds that would otherwise remain hidden within the noise of normal system operations. Taken together, these mitigations for the Design and Setup phase seek to reduce the probability of initial compromise through user training and technical controls, constrain the misuse of valid accounts through identity management, protect credentials and infrastructure through encryption and system hardening, preserve the integrity of software and AI components through supply chain verification, and detect persistent threats through correlation of events over time. Having established these foundational defenses, we now examine the mitigations that protect the model development and evaluation phase.

\subsection{Mitigations for Model Development and Evaluation} \label{sec:model_development_and_evaluation_mitigations}
Mitigations for the \textit{Model Development and Evaluation} phase target the integrity and confidentiality of data, the robustness of training and validation pipelines, and the trustworthiness of model artifacts. Fig.~\ref{fig:mitre-defenses} provides a high level overview of the main defenses, and this subsection examines them in more detail by first considering data-focused mitigations, then detection and validation mechanisms, and finally model-focused measures. These complementary strategies work together to ensure that both the data feeding into models and the models themselves remain trustworthy throughout the development lifecycle.

Data-focused defenses aim to keep poisoned, redundant, or sensitive records out of training sets while preserving data utility for learning, requiring a careful balance between security and model performance. Sanitizing training data~\cite{SanitizeTrainingData} identifies and removes compromised inputs before they enter the pipeline, using techniques such as anomaly detection, statistical outlier removal, and consistency checking to filter malicious samples. Data and model lineage~\cite{ModelDistributionMethods} provides visibility into data origins and changes and enables comparison of current datasets with trusted baselines, creating an auditable trail that supports forensic analysis when poisoning is suspected. Broader work on adversarial ML mitigations emphasizes deduplication and sanitization to remove redundant or sensitive samples~\cite{esmradi2023comprehensive}. Encrypting data and models~\cite{esmradi2023comprehensive} secures training data, gradients, and model outputs from disclosure, with differential privacy injecting carefully calibrated noise to protect private information during training~\cite{zhang2018mitigating} and encryption-based approaches preventing unauthorized access to sensitive model parameters. Data anonymization and end-to-end encryption further protect personal information in training corpora~\cite{esmradi2023comprehensive}. To limit resource abuse during training, proof-of-work puzzles can slow extraction attempts~\cite{koizumi2015study}, and profiling natural examples with energy consumption thresholds can help detect attacks that maximize resource usage to either degrade service availability or facilitate model extraction through repeated queries~\cite{jaramillo2018malware}. Limiting the release of information further constrains what attackers can learn about model internals, training data, or system architecture through both passive observation and active probing. Together, these data-focused mitigations create multiple barriers that protect training data from poisoning, preserve privacy through encryption and differential privacy, and limit the information available to attackers through careful control of what data and model details are exposed.

Complementing these preventive measures, detection and validation mechanisms provide ongoing assurance that data quality and model integrity are maintained throughout development, enabling teams to identify and respond to compromises before they propagate into production. Vulnerability scanning~\cite{VulnerabilityScanning} can uncover weaknesses in model artifacts and in supporting software such as pickle file loaders, which are often used to serialize trained models but can execute arbitrary code when loading malicious files. An enhanced machine learning-based detection scheme for SCADA networks, for instance, achieved 99.49\% accuracy, 99.23\% precision, and 99.75\% recall when identifying attacks~\cite{Ahakonye2021}. Sanitizing training data remains important throughout the lifecycle, not just at the initial ingestion stage. In Anti Backdoor Learning, Li et al.~\cite{antibackdoor} extend anomaly detection during training to remove both attack samples and non-regular traffic, leading to a more accurate detector with high detection rates and low false positives~\cite{antibackdoor}. Regular validations~\cite{Zhao2025} can reveal backdoors, adversarial bias, or data poisoning by testing for concept drift and shifts in training data distributions. One study proposed a mathematical framework for distribution shift and introduced a tangent space regularized estimator to control the shift and improve long-term accuracy~\cite{Zhao2025}. Verifying artifacts \cite{VerifyMLArtifacts} through cryptographic checksums ensures that datasets and models have not been tampered with during storage or transfer, providing a technical guarantee that complements behavioral detection methods. Maintaining a complete record of dataset sources and modifications supports these validation checks and improves transparency for later audits, enabling security teams to reconstruct the provenance of any suspicious artifacts and determine whether they originated from trusted sources or may have been compromised. These detection and validation mechanisms work in concert to provide continuous assurance, catching both subtle poisoning attempts that evade initial sanitization and supply chain attacks that introduce compromised artifacts at later stages of development.

While data quality and validation form the foundation, model-centric mitigations focus on robustness and integrity to ensure that the models themselves resist manipulation and perform reliably even under adversarial conditions. Adversarial training~\cite{Ma2023} strengthens models by exposing them to adversarial examples during the training process, teaching them to maintain correct predictions even when inputs are perturbed. Recent Increasing Margin Adversarial training improves on prior methods by gradually enlarging classification margins and has been shown to outperform other defenses on six public image datasets for both classification and segmentation tasks~\cite{Ma2023}. Model hardening~\cite{ModelHardening} encompasses a broader set of techniques including defensive distillation, gradient masking, and input preprocessing that strengthen models against manipulations or filter unusual inputs before inference~\cite{balantrapu2019adversarial}. Randomized smoothing and constraints such as Lipschitz continuity can yield certified robustness guarantees within defined perturbation thresholds~\cite{balantrapu2019adversarial}, providing mathematical assurance that models will maintain correct behavior for inputs within a bounded distance from training examples. Detection tools build on anomaly detection or auxiliary models to flag suspicious samples, while explainable mitigation strategies help teams understand and respond to new attack patterns~\cite{balantrapu2019adversarial}. Using ensemble methods~\cite{UseEnsembleMethods} can increase robustness against adversarial inputs by combining heterogeneous models that make errors on different samples, making it harder for attackers to find inputs that fool all models simultaneously. In vulnerability classification, a stacking learning classifier with heterogeneous ensemble methods has been shown to be more robust, reliable, and accurate than single models~\cite{Wang2020}. These model-centric mitigations create models that are inherently more resistant to adversarial manipulation, reducing the likelihood that attackers can exploit them through carefully crafted inputs or extract sensitive information through model inversion attacks. These mitigations for the Model Development and Evaluation phase therefore address data quality through sanitization and encryption, provide continuous validation through scanning and drift detection, and build robust models through adversarial training and ensemble methods. They aim to ensure that training data remains trustworthy, that models are evaluated under realistic adversarial conditions, and that all artifacts can be verified throughout the development lifecycle. With development phase protections in place, we next examine how these defenses extend into the Operations phase where models face real world threats.

\subsection{Mitigations for Operations} \label{sec:operations_mitigations}
Mitigations in the \textit{Operations} phase address attacks that occur when models are deployed and exposed to users, services, and networks. Fig.~\ref{fig:mitre-defenses} situates these defenses within the overall taxonomy, and this subsection discusses how they restrict access to deployed systems, provide telemetry for detection and response, harden inference against adversarial use, and address challenges specific to large language models. The operational security posture builds upon the foundations established in earlier phases while introducing runtime-specific protections that defend against attacks targeting deployed models.

A first group of mitigations restricts who can reach deployed models and what they can do once they have access, creating multiple layers of access control that constrain both external attackers and insider threats. Organizations can enforce strong authentication for production APIs and control model access~\cite{ControlAccesstoMLModelsandDataatRest} by restricting ML model and data access in production environments. One practical guidance set recommends strong access controls for LLM hosting platforms, code repositories, and training environments and stresses restricting LLM access to network resources, internal services, and APIs~\cite{Securityplanningfor}. An Agentic AI Identity Management approach similarly emphasizes continuous verification for real time authentication, least privilege access, and micro segmentation to limit lateral movement~\cite{agenticaiidentity}. Restricting model queries~\cite{RestrictNumberofMLModelQueries} protects against denial of ML service and excessive probing by limiting the number and rate of queries allowed for each model. At the network interface level, rate limiting and input validation and sanitization~\cite{threatmodeling} on all proposed queries gate users from the actual model and reduce the chance of exploiting application logic or overwhelming the system with malicious requests. Network segmentation~\cite{NetworkSegmentation_mitre} further isolates production ML systems from other parts of the infrastructure, preventing attackers who compromise one system from easily pivoting to attack models. Filtering network traffic~\cite{FilterNetworkTraffic_mitre} blocks malicious communications and prevents data exfiltration by inspecting packets at network boundaries and enforcing policies about which services can communicate with deployed models. These access control and network security measures work together to create a defense-in-depth posture where attackers must overcome multiple barriers to reach deployed models, and where each successful bypass still leaves them constrained by remaining controls.

Beyond access restrictions, monitoring and telemetry provide visibility into how deployed models are used and enable rapid detection of anomalous behavior that may indicate attacks or system failures. AI telemetry logging~\cite{AITelemetryLogging} and monitoring model queries in real time enables detection of suspicious patterns, potential exfiltration attempts, and misuse of privileged operations. An intelligent observability framework that combines AI telemetry and infrastructure metrics in a Kubernetes testbed reported a 35\% improvement in anomaly detection accuracy and a reduction of mean time to resolution by more than 40\% compared with baseline tools~\cite{Nimmagadda2025}. Traditional network security controls complement AI-specific monitoring. Network intrusion detection~\cite{NetworkIntrusionPrevention_mitre} systems, web application firewalls, and TLS-based secure communication for public-facing applications help prevent exploitation attempts. Additional practices such as input validation and sanitization, least privilege access to back end services, application isolation and sandboxing, exploit protection, network segmentation, and privileged account management provide multiple layers of control and visibility~\cite{Ding_2024,mitreatlas}. Frequent vulnerability scans and timely software updates~\cite{UpdateSoftware_mitre} further reduce exposure by identifying and patching weaknesses before they can be exploited. A case study on deep learning-based intrusion detection shows that deep neural networks can outperform other machine learning approaches for identifying network-based attacks~\cite{acasestudy}. These monitoring and traditional security controls work synergistically, with AI telemetry providing visibility into model-specific behaviors while network monitoring and intrusion detection catch broader infrastructure attacks, together creating comprehensive visibility across the entire production environment.

\begin{table*}[!htbp]
    \caption{Summary of MLOps mitigation strategies with empirical validation. Traditional IT column indicates defenses applicable to conventional systems (\ding{51}) or is unique to ML environments (\ding{53}). The empirical validation column cites industry implementations and case studies demonstrating practical deployment and effectiveness of each technique.}
    \begin{adjustbox}{max width=\textwidth}
        \begin{tabular}{|p{2cm}|p{4.8cm}|p{5cm}|p{2cm}|}
\hline
\rowcolor[HTML]{E1FFFF} 
{\color[HTML]{010101} \textbf{Taxonomy}} & {\color[HTML]{010101} \textbf{Mitigation Techniques}} & {\color[HTML]{010101} \textbf{Empirical Validation}} & {\color[HTML]{010101} \textbf{Traditional IT}} \\ \hline

\multirow{12}{2cm}{Design and Setup} 
 & User Training \cite{UserTraining} & Security Awareness Training (SAT) \cite{SecurityAwarenessTraining} & \ding{51} \\ \cline{2-4} 
 & Strong Password Policies \cite{PasswordPolicies_mitre} & Global Info Study \cite{globinfo} & \ding{51} \\ \cline{2-4} 
 & Multi-factor Authentication \cite{Multi-factorAuthentication_mitre} & Azure AD Report \cite{Meyer2023} & \ding{51} \\ \cline{2-4} 
 & Regular Audits \cite{audit_mitre} & Indonesia IT Audits \cite{inplimentationofinformation} & \ding{51} \\ \cline{2-4}
 & Intrusion Prevention System \cite{BehaviorPreventiononEndpoint_mitre} & Cisco case study \cite{Reid1992} & \ding{51} \\ \cline{2-4} 
 & Privilege Account Management \cite{PrivilegedAccountManagement_mitre} & Segura Privileged Access \cite{completeguide} & \ding{51} \\ \cline{2-4} 
 & Secure Key Management & Google Cloud Key Management \cite{cloudkeymanagement} & \ding{51} \\ \cline{2-4} 
 & Encrypt Sensitive Data \cite{EncryptSensitiveInformation_mitre} & Thales Encryption \cite{theenterprise} & \ding{51} \\ \cline{2-4} 
 & Software Bill of Materials \cite{securingthesw} & Forbes \cite{ForbesSoftwareBillofMaterials} & \ding{51} \\ \cline{2-4} 
 & Code Signing \cite{CodeSigning} & AWS Lambda Signing \cite{bestpracticesandadvanced}  & \ding{53} \\ \cline{2-4} 
 & AI Bill of Materials \cite{AIBillofMaterials} & Wiz AI-BOM \cite{WizAIBillofMaterial} & \ding{53} \\ \cline{2-4} 
 & Cyber Persistence Detector \cite{liu2024accurate} & CrowdStrike Detector \cite{CrowdStrikeAPTDetector}  & \ding{51} \\ \hline

\multirow{10}{2cm}{Model Development and Evaluations} 
 & Sanitize Training Data \cite{SanitizeTrainingData} & AWS Data Cleansing \cite{AWSDataCleansing} & \ding{53} \\ \cline{2-4} 
 & Data and Model Lineage \cite{ModelDistributionMethods} & AWS Data and Model Lineage \cite{AWSDataAndModelLineage} & \ding{53} \\ \cline{2-4} 
 & Verify Artifacts \cite{VerifyMLArtifacts} & Netflix \cite{MLOpsAdoption} & \ding{53} \\ \cline{2-4} 
 & Encrypt Data and Model \cite{esmradi2023comprehensive} & AWS Sagemaker \cite{aws_fhe_sagemaker_2023} & \ding{51} \\ \cline{2-4} 
 & Vulnerability Scanner \cite{VulnerabilityScanning} & ModelScan \cite{protectai_modelscan_2025} & \ding{53} \\ \cline{2-4} 
 & Limit Release of Information \cite{LimitPublicReleaseofInformation} & \ding{53} & \ding{53} \\ \cline{2-4} 
 & Regular Validations \cite{Zhao2025} & IBM Watson OpenScale \cite{ibm_manage_model_risk_2024} & \ding{53} \\ \cline{2-4} 
 & Adversarial Training \cite{Ma2023} & Wiz AI-SPM \cite{wiz_adversarial_ai_2024} &  \ding{53} \\ \cline{2-4} 
 & Use Ensemble Methods \cite{UseEnsembleMethods} & IBM \cite{ibm_ensemble_learning_2023}  & \ding{53} \\ \cline{2-4}
 & Model Hardening \cite{ModelHardening} & Avaly \cite{avaly_robust_model_hardening_2025}  & \ding{53} \\ \hline

\multirow{13}{2cm}{Operations} 
 & Adversarial Input Detection \cite{AdversarialInputDetection} & AWS \cite{aws_detect_adversarial_inputs_2022} & \ding{53} \\ \cline{2-4} 
 & Input Validation and Sanitization \cite{threatmodeling} & Wiz \cite{wiz_llm_security_2025} & \ding{53} \\ \cline{2-4} 
 & Restrict Model Queries \cite{RestrictNumberofMLModelQueries} & Microsoft Azure ML \cite{microsoft_azureml_enterprise_security_2025} & \ding{53} \\ \cline{2-4} 
 & Control Model Access \cite{ControlAccesstoMLModelsandDataatRest} & Microsoft Azure ML \cite{microsoft_azureml_enterprise_security_2025} & \ding{53} \\ \cline{2-4} 
 & Multi-Modal Sensors \cite{UseMulti-ModalSensors} & AWS \cite{aws_physical_world_ai_2025} & \ding{53} \\ \cline{2-4} 
 & AI Telemetry Logging \cite{AITelemetryLogging} & IBM Instana \cite{ibm_instana_genai_observability_2024} & \ding{53} \\ \cline{2-4} 
 & LLM Tagging \cite{lee2024prompt} & Appen \cite{appen_data_annotation_2025} & \ding{53} \\ \cline{2-4} 
 & Generative AI Guidelines \cite{GenerativeAIGuidelines} & Google \cite{google_gemini_api_safety_guidance_2025} & \ding{53} \\ \cline{2-4} 
 & Generative AI Guardrails \cite{GenerativeAIGuardrails} & AWS Bedrock Guardrails \cite{aws_bedrock_guardrails_2025} & \ding{53} \\ \cline{2-4} 
 & Network Intrusion Detection \cite{NetworkIntrusionPrevention_mitre} & Google Cloud IDS \cite{google_cloud_ids_2025} & \ding{51} \\ \cline{2-4} 
 & Network Segmentation \cite{NetworkSegmentation_mitre} & Microsoft Azure ML \cite{microsoft_azureml_network_isolation_2025}  & \ding{51} \\ \cline{2-4} 
 & Filter Network \cite{FilterNetworkTraffic_mitre} & Microsoft \cite{microsoft_vnet_service_endpoints_2025} & \ding{51} \\ \cline{2-4} 
 & Software Updates \cite{UpdateSoftware_mitre} & Apple Security Bounty \cite{apple_security_bounty_2025} & \ding{51} \\ \hline
\end{tabular}
    \end{adjustbox}
    \label{tab:defenses-mapping}
\end{table*}

At the inference level, specialized defenses focus on the content of queries and responses to prevent adversarial manipulation of model behavior and protect against attacks that exploit model vulnerabilities through carefully crafted inputs. Adversarial input detection~\cite{AdversarialInputDetection} evaluates incoming requests against known legitimate patterns and filters likely adversarial inputs before they reach the model, using statistical tests, learned classifiers, or consistency checks to identify suspicious queries. Model hardening techniques aim to make models less sensitive to small perturbations and more robust to adversarial attacks. In one benchmark, adversarial training for a deep neural network-based intrusion detection system improved robustness, achieving F1 scores of 93\%, 99\%, 85\%, and 83\% for benign, backdoor, ransomware, and cross-site scripting classes respectively~\cite{improvingtherobustness}. Multi-modal sensors~\cite{UseMulti-ModalSensors} can reduce risks in physical environments by combining diverse data sources to detect unauthorized access and tampering, particularly important for embodied AI systems such as autonomous vehicles or robotic systems that interact with the physical world. Pre-processing and input restoration techniques help counter physical perturbations such as adversarial stickers or projections that attempt to fool vision systems through modifications to the physical environment rather than digital inputs. These inference-level defenses create multiple checkpoints where adversarial inputs can be detected and rejected, hardened models that maintain correct behavior even when attacks succeed, and specialized protections for physical world scenarios where digital defenses alone are insufficient.

LLM deployments introduce unique security challenges that require additional specialized controls beyond traditional ML systems, particularly around prompt injection, jailbreaking, and the generation of harmful content. Generative AI guardrails~\cite{GenerativeAIGuardrails} sit between user interactions and model outputs and can implement filters, rule-based logic, or classifier-based checks to enforce safety constraints. A recent control-theoretic guardrail approach goes beyond refusal and actively corrects risky outputs to safe ones, as demonstrated in simulated driving and e-commerce experiments~\cite{Pandya2025}. Generative AI guidelines~\cite{GenerativeAIGuidelines} define acceptable model behavior and are often embedded into system prompts, establishing boundaries around topics the model should not discuss or actions it should not take. LLM tagging~\cite{lee2024prompt} marks model outputs so that downstream components can identify their origin and distinguish between user-provided content and model-generated responses. Delimiting data and random sequence enclosure make it harder for prompts to be copied across interactions~\cite{lee2024prompt}, while the Sandwich Defense layers user instructions around previous responses, and Instruction Defense clarifies that models must not alter user inputs~\cite{lee2024prompt}. Marking prompts and outputs~\cite{lee2024prompt} further distinguishes user content from agent-generated content, helping to prevent prompt injection attacks where malicious instructions are hidden within seemingly benign inputs. Combined with guardrails and guidelines, these techniques reduce cross-model injection attempts and limit the spread of malicious prompts, although they also raise new security considerations if higher capability models that enforce the guardrails become compromised. Regular patching, developer and user education, thorough logging, and frequent audits close additional attack surfaces that attackers might otherwise exploit~\cite{branescu2024automated}, while alignment techniques such as supervised fine-tuning and reinforcement learning from human feedback keep models aligned with organizational policy over time. These LLM-specific defenses address the unique attack surface created by natural language interfaces, where the boundary between data and instructions is fundamentally blurred and where models must distinguish between legitimate user requests and attempts to manipulate their behavior through carefully crafted prompts.

Across these three subsections we have outlined mitigation strategies that span the design and setup phase, the model development and evaluation phase, and the operations phase. Together, these defenses aim to protect MLOps and LLMOps deployments from user compromise and data gathering attacks through to model training, deployment, and runtime abuse. Table~\ref{tab:defenses-mapping} summarizes the mitigation strategies discussed, their empirical validation through industry implementations, and their applicability beyond ML systems. However, the current mitigations remain fragmented and often reactive, and many combinations of tools and practices are not yet well understood in complex real world pipelines. In the next section we therefore turn to research recommendations, highlighting open challenges and directions for future work that can strengthen MLOps security across these interconnected domains and support more systematic lifecycle-aware defenses.

\section{Research Challenges and Recommendation} \label{sec:research-challenges}
Although the previous section outlined a range of mitigation strategies to defend the MLOps ecosystem, these measures often provide only partial protection against increasingly sophisticated, AI-driven adversaries. As attacks evolve rapidly, organizations must address not only technical vulnerabilities but also legal, ethical, and operational complexities. To address RQ4, this section identifies notable gaps that remain in current practices and proposes future research directions needed to strengthen security across the MLOps ecosystem. The following discussion highlights key concerns and proposes future research directions, with the goal of strengthening the security of the MLOps ecosystem while preserving essential principles of transparency, privacy, and compliance.

\begin{enumerate}
    \item \textbf{The Rise of AI-Driven Social Engineering:} Attackers now exploit generative AI to craft highly convincing phishing emails, clone voices for vishing schemes, and produce deepfake videos that entice employees into revealing sensitive credentials \cite{google-ai-vishing-attacks}. Once inside an organization, these adversaries can move laterally and compromise vital MLOps components. Although zero trust models and ongoing user education are vital, they frequently fall behind AI's rapid advancement. Further research should refine deepfake detection techniques, enhance digital authentication protocols, and implement comprehensive incident monitoring to counter the increasingly realistic nature of AI-enabled social engineering.
    \item \textbf{Malicious Repositories and LLM Hallucinations:} LLMs occasionally generate misleading ``hallucinations'', which can enable attackers to insert malicious packages into trusted online repositories. If organizations inadvertently integrate these compromised resources, they risk exposing intellectual property, disrupting workflows, or introducing harmful payloads. Threat actors may also upload cloned or tampered models to platforms such as Hugging Face, deceiving unsuspecting practitioners into deploying compromised versions \cite{chatgpt_hallucination}. Future research should explore automated systems for detecting suspicious content, robust versioning protocols to ensure code integrity, and community-driven validation frameworks supported by the broader MLOps ecosystem.
    \item \textbf{Misuse of AI Tools by Advanced Persistent Threats (APTs):} Advanced Persistent Threats (APTs) groups sponsored by nation state actors increasingly rely on freely available powerful AI tools, such as Google's Gemini, to identify vulnerabilities, develop sophisticated malware, and evade standard security measures \cite{google-misuse-gemini}. Google DeepMind's Gemma team has raised concerns about potential misuse of AI in upcoming releases \cite{gemmateam2024gemmaopenmodelsbased}. Researchers should examine subtle ways AI may be weaponized, establish precise ethical and legal standards, and propose robust prevention strategies. Effectively defending MLOps ecosystem against increasingly sophisticated adversaries will require detecting and preventing abuse of emerging AI technologies.
    \item \textbf{Benchmarking MLOps Defenses:} Many firms rely on laboratory-style tests that do not accurately reflect real-world attacks. As a result, MLOps ecosystem remain vulnerable when deployed in practical settings. Future research should focus on benchmarking defenses against adaptive and diverse attack patterns \cite{NIST2024}. By simulating adversarial scenarios, such as targeted phishing attempts and sophisticated AI-powered attacks, MLOps teams can evaluate how well their security measures perform under pressure. Additionally, continuous research should establish standardized criteria for comparing defensive mechanisms. This approach will support ongoing improvement and ensure better alignment with real operational risks.
    \item \textbf{Red-Teaming for Robust Security:} Red team exercises serve as an effective approach for uncovering hidden vulnerabilities, validating existing defenses, and strengthening overall system resilience. Drawing on initiatives such as MITRE’s red team program, organizations can more accurately simulate real-world attack scenarios and identify exploit paths that may remain undetected during conventional testing. Future research should explore frameworks that support recurring, collaborative red-teaming efforts across sectors, ensuring that shared insights contribute to a more secure global MLOps ecosystem. By embracing these rigorous assessments, security teams can enhance threat detection, improve incident response, and cultivate a proactive security posture.
    \item \textbf{Strengthening MLOps Robustness:} A robust MLOps strategy adopts security as a default posture and embeds high-assurance practices throughout the ML lifecycle. Regular testing using established adversarial simulation tools and tailored threat scenarios can help uncover and remediate vulnerabilities before they lead to critical failures. Further investigation is needed to identify optimal strategies for streamlining configuration management, continuous patching, and resilience assessments across the MLOps ecosystem. This proactive approach enables organizations to anticipate, disrupt, and adapt to evolving adversarial threats.
    \item \textbf{Open Source vs. Closed Source:} An important question in securing MLOps is how to balance openness with security when comparing closed-source and open-source approaches. Publicly accessible code fosters collaborative innovation but may also provide adversaries with insights into system architecture. Conversely, limited disclosure can hinder attackers but may reduce transparency and erode user trust. Further research should explore how community-driven governance, selective disclosure, and cryptographic verification can preserve the benefits of open development while reducing reconnaissance risks. By striking this balance, organizations can maintain stakeholder trust and safeguard operational integrity.
    \item \textbf{Cultivating Cyber-Hygiene Among MLOps Practitioners:} Developers, DevOps engineers, and data scientists often work in silos, unintentionally introducing security risks in MLOps environments. While password encryption and secure networking are crucial, practitioners also need training in secure coding \cite{UserTraining}, proactive vulnerability management, and regular data backups. The compelling and clear evidence for this is that Effective security awareness training, with regular simulated phishing exercises, educates employees and significantly reduces the human risk of cybersecurity threats \cite{SecurityAwarenessTraining}. Further research is needed to assess the effectiveness of training methods, incentive structures, and policies in fostering a security-oriented culture at both organizational and societal levels.
    \item \textbf{Balancing Transparency with Trust:} Organizations often publish white papers and technical reports to showcase their commitment to ethical and responsible AI. However, excessive disclosure of MLOps infrastructure can provide adversaries with a blueprint for exploitation. Researchers must develop frameworks that promote transparency in fairness criteria and privacy protections without revealing sensitive operational details. Future research should focus on defining indicators of responsible AI that offer public and regulatory assurance without compromising system security. A well-balanced transparency model can help organizations build trust while safeguarding confidential information.
    \item \textbf{The Challenge of Copyright, Privacy, and Compliance:} Generative AI models may unintentionally expose copyrighted or sensitive personal data, leading to significant legal consequences and reputational damage. When models continuously train on real or user-provided data, issuing disclaimers provides only a minimal level of protection \cite{bloomberg-chatgpt}. Scholars should develop comprehensive solutions that include frequent model audits, automated data validation, and tightly controlled input and output streams. These measures can help AI advancements comply with legal requirements while safeguarding user privacy and intellectual property. Recent work on model watermarking suggests that such techniques can contribute to post hoc copyright enforcement by enabling the detection of stolen models, yet they remain a passive defense whose current designs are mostly evaluated on small scale image classification tasks, leaving open questions about scalability and applicability to other machine learning domains before they can be broadly adopted in practice \cite{Franziska_watermark}. Future work should strengthen the stability of watermarking schemes, integrate them with active security controls, extend them beyond small image classification settings to diverse data types and learning tasks, and align their use with real world legal and organizational processes for demonstrating and enforcing ownership claims \cite{Franziska_watermark}.

\end{enumerate}
    Protecting MLOps ecosystem requires a holistic strategy that combines technological solutions with organizational best practices. By embracing rigorous benchmarking, conducting thorough red-team assessments, and establishing frameworks that prioritize security from the onset, organizations can better defend their MLOps ecosystems against emerging attacks. 

\section{Conclusion}
In recent years, MLOps has fundamentally reshaped how machine learning models are developed, deployed, and maintained. It offers substantial advantages, including enhanced scalability, reproducibility, rapid deployment, and improved team collaboration. However, the acceleration of development to deployment cycles, combined with increased automation and highly interconnected workflows, significantly expands the attack surface. This heightens the risk of vulnerabilities and security breaches. As machine learning becomes increasingly embedded in critical systems, these trends underscore the urgent need for cybersecurity frameworks that can evolve alongside technological advancements.

This study highlights that while MLOps delivers critical operational benefits, it also introduces security challenges that demand focused attention. Leveraging the MITRE ATLAS framework for AI-specific threats, we present a structured taxonomy of vulnerabilities spanning the entire MLOps lifecycle, from data collection to deployment. Insights from red-team exercises and real-world incidents illustrate how adversaries exploit both user-level and system-level weaknesses. To address these threats, we propose targeted mitigation strategies informed by established frameworks and best practices. These recommendations offer guidance for improving MLOps security and identifying priorities for future research.

As MLOps adoption continues to grow, organizations must embrace a proactive security posture by embedding robust practices early in their machine learning workflows. By implementing the frameworks and recommendations outlined in this work, they can more effectively safeguard models, maintain stakeholder and customer confidence, and ensure that machine learning continues to support secure, ethical, and responsible innovation. In doing so, they establish a resilient foundation for long-term security, sustained innovation, and adaptive growth in a rapidly evolving threat landscape.

\section*{Acknowledgment}
This work was supported by the Predictive Analytics and Technology Integration (PATENT) Laboratory. The author gratefully acknowledges the valuable feedback, discussions, and support provided by the laboratory members, which contributed to improving the quality of this paper.


\bibliographystyle{acm}
\bibliography{sources,sources_web}

\end{document}